\documentclass[%
 reprint,
 amsmath,amssymb,
 aps,prl
]{revtex4-1}

\usepackage[T1]{fontenc}
\usepackage{amsfonts,amssymb}
\usepackage{color}
\usepackage{graphicx}
\usepackage{dcolumn}
\usepackage{bm}
\usepackage{amsmath}
\usepackage{latexsym,amscd}
\usepackage{epstopdf}
\usepackage{epsfig}
\usepackage{wrapfig}
\usepackage{txfonts}
\usepackage[export]{adjustbox}

\begin{document}

\preprint{APS/123-QED}

\title{Quantum-correlated photons from semiconductor cavity polaritons}

\author{Guillermo Mu\~noz-Matutano, Andrew Wood, Mattias Johnson, Xavier Vidal Asensio, Ben Baragiola, Andreas Reinhard, Thomas Volz}
\email{thomas.volz@mq.edu.au, guillermo.munozmatutano@mq.edu.au}
 \affiliation{Department of Physics and Astronomy, Macquarie University, Sydney, New South Wales, Australia.}
 \affiliation{ARC Centre of Excellence for Engineered Quantum Systems, Macquarie University, NSW 2109, Australia\\}

\author{Aristide Lemaitre, Jaqueline Bloch, Alberto Amo}
 \affiliation{Centre de Nanosciences et de Nanotechnologies, CNRS, Univ. Paris-Sud, Universit\`{e} Paris-Saclay, C2N-Marcoussis, 91460 Marcoussis, France\\}
 
\author{Benjamin Besga}
\affiliation{Institut N\'{e}el, Universit\'{e} Grenoble Alpes - CNRS: UPR2940, 38042 Grenoble, France}
\affiliation{Department of Physics and Astronomy, Macquarie University, Sydney, New South Wales, Australia.}
\affiliation{ARC Centre of Excellence for Engineered Quantum Systems, Macquarie University, NSW 2109, Australia\\}

\author{Maxime Richard}
\affiliation{Institut N\'{e}el, Universit\'{e} Grenoble Alpes - CNRS: UPR2940, 38042 Grenoble, France\\}

\begin{abstract}

\end{abstract}

\pacs{Valid PACS appear here}
\maketitle

\textbf{Over the past decade, exciton-polaritons in semiconductor microcavities have attracted a great deal of interest as a driven-dissipative quantum fluid \cite{Carusotto2013}. These systems offer themselves as a versatile platform for performing Hamiltonian simulations with light \cite{Berloff2017,Jacqmin2014,Baboux2016}, as well as for experimentally realizing nontrivial out-of-equilibrium phase transitions \cite{Dagvadorj2015}. In addition, polaritons exhibit a sizeable mutual interaction strength that opens up a whole range of possibilities in the context of quantum state generation. While squeezed light emission from polaritons has been reported previously \cite{Bramati2014}, the granular nature of polaritons has not been observed to date \cite{Sanvitto2016}. The latter capability is particularly attractive for realizing strongly correlated many-body quantum states of light on scalable arrays of coupled cavities \cite{Angelakis2017}. Here we demonstrate that by optically confining polaritons to a very small effective mode volume, one can reach the {\it weak blockade regime}, in which the nonlinearity turns strong enough to become significant at the few particle level, and thus produce a non-negligible antibunching in the emitted photons statistics \cite{Verger2006}. Our results act as a door opener for accessing the newly emerging field of quantum polaritonics, and as a proof of principle that optically confined exciton-polaritons can be considered as a realistic, new strategy to generate single photons. }

\begin{figure*}
\begin{center}
\includegraphics[scale=0.45]{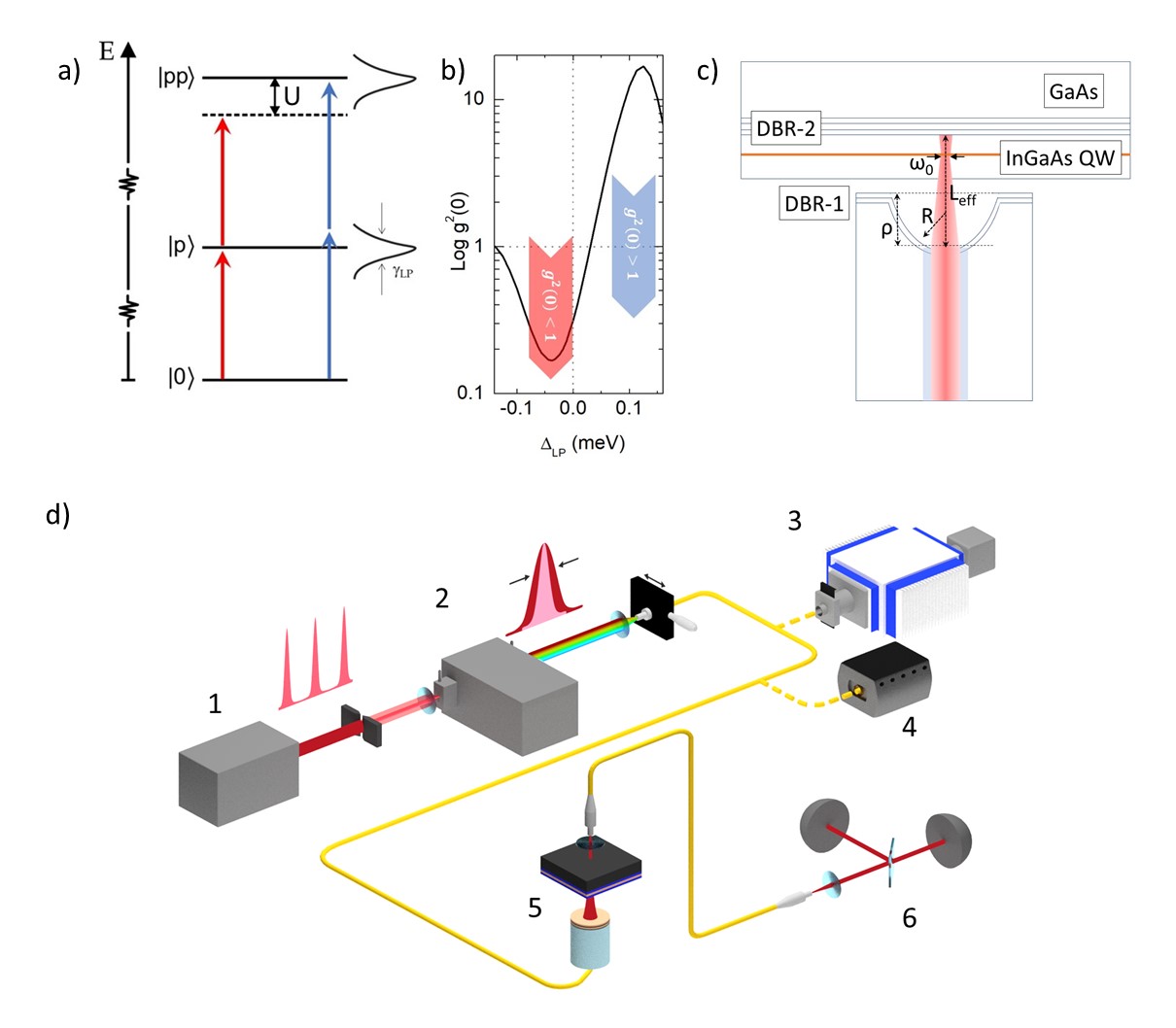}
\caption{\textbf{Polariton Blockade and Setup}. a) Single-polariton excitation spectrum. Polariton-polariton interactions shift the two-polariton state to higher energies. If the shift $U$ is larger than the polariton linewidth $\gamma_{LP}$, polariton blockade occurs: A photon resonant with (or slightly red-detuned from) the transition from the ground state to the one-polariton state will generate a polariton, but a second photon of the same colour cannot excite the system further. It gets blocked by the first photon. For blue-detuned photons, the chance of two photons entering the system simultaneously is enhanced. As a result, photons are expected to exhibit antibunching for red detuning and bunching for blue detuning, as depicted in b) which shows a numerical prediction of $g^{(2)}(0)$ as function of laser detuning for the case $U \gg \hbar\gamma_{LP}$. c) Basic fiber-cavity layout consisting of the fiber-tip with dielectric Bragg reflectors (DBR-1), the GaAs cavity spacer layer, the InGaAs quantum well (QW) on top of a stack of alternating layers of AlAs/GaAs (DBR-2). The system parameters are the fiber curvature radius (R), the mirror depth ($\rho$), the effective cavity length ($L_{eff}$) and cavity mode waist ($\omega_0$). d) Experimental setup. The key ingredients for the experiment are the resonant pulsed excitation (1), laser pulse shaping (2), laser pulse monitoring with a streak camera (3), wavelength read-out using a wavemeter (4), the fiber cavity microscope at 4K (5) and the standard free-space Hanbury-Brown and Twiss interferometer for correlation measurements (6).} 
\label{fig_1} 
\end{center}
\end{figure*}

Semiconductor polaritons are half-matter half-light quasiparticles that form when an elementary excitation such as a quantum-well exciton interacts sufficiently strongly with light. Signature of polariton formation is the Rabi splitting at resonance, resulting in a lower polariton (LP) and an upper polariton (UP) branch. Within the rotating frame approximation, the generic 2D exciton-polariton quasiparticle is described by bosonic excitations of a harmonic oscillator. At higher quasiparticle densities this description breaks down due to polariton-polariton interactions, leading to a Kerr-like nonlinearity that can be exploited to realize parametric down-conversion \cite{Savasta2005}, squeezing \cite{Bramati2014} and optical spin switches \cite{Amo2010}. This nonlinearity can be enhanced by spatial confinement of the polariton wavefunction: the smaller the confinement the bigger the polariton-polariton interaction. When the nonlinearity turns large enough, correlations build up at the few-particle level. In particular, under resonant optical excitation, the statistics of the transmitted light starts to exhibit pure quantum behavior. For laser light resonant (or slightly red-detuned) with the fundamental polariton transition, the scattered photons exhibit antibunching, i.e. the corresponding auto-correlation function at zero time-delay, $g^{(2)}(0)$ drops below 1 (see figure \ref{fig_1}.a-b). This behavior is reminiscent of the photon blockade effect in cavity QED \cite{Imamoglu1997} and to Rydberg blockade effect in atomic physics \cite{Liebisch2005}. For polaritons, the corresponding effect was termed {\it polariton blockade} \cite{Verger2006}. Recent experiments based on Rydberg polaritons \cite{Ningyuan2017} have reported dissipative blockade \cite{Carusotto2010}. The ease of use and the potential for on-chip integration, however, make semiconductor cavity polaritons the prime platform for translating strongly interacting photonic quantum systems into real-world applications. Demonstrating quantum correlations between polaritons is a key step towards this long-term goal. 

\begin{figure*}
\begin{center}
\includegraphics[scale=0.45]{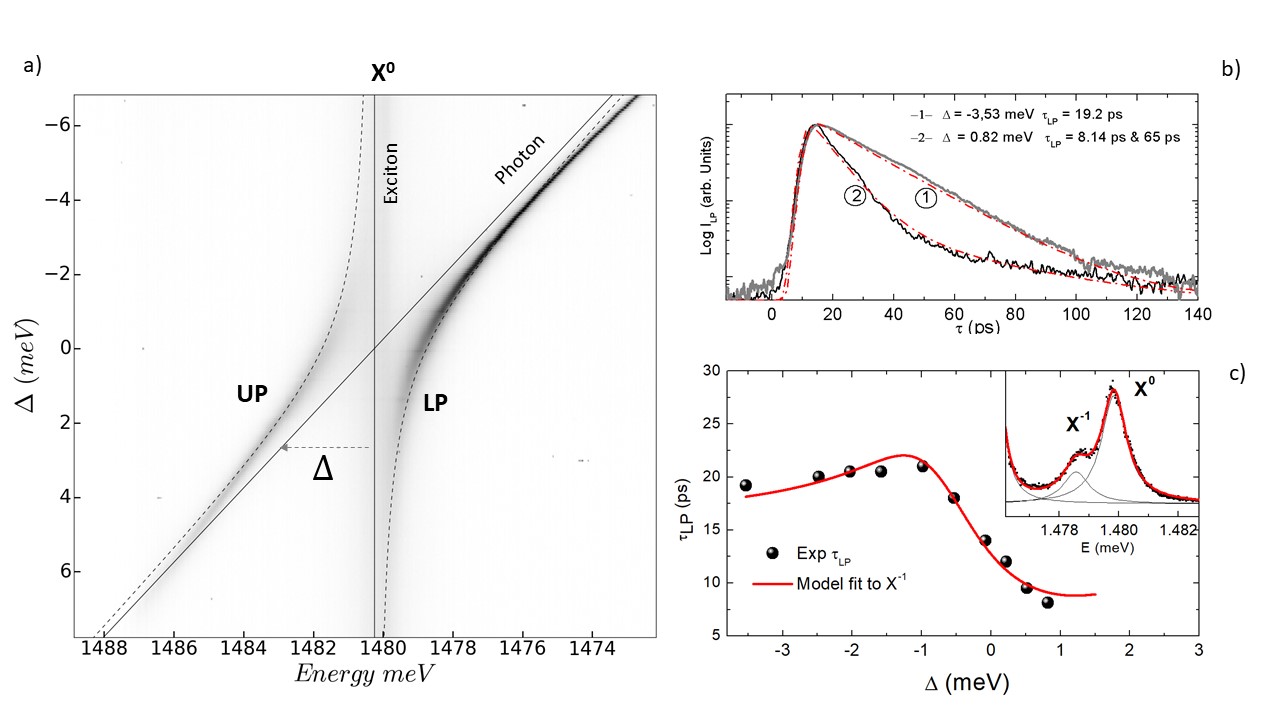} 
\centering
\caption{\textbf{Avoided crossing and polariton lifetimes}. a) Avoided crossing of the bare exciton and the cavity mode, forming Lower (LP) and Upper (UP) Polariton dressed states near resonance ($\Delta \approx$ 0). Dotted lines correspond to fit based on a two-coupled oscillator model, yielding a Rabi Vacuum splitting of $\Omega$ = 2g = 3 meV. b) Experimental LP lifetime decay curve recorded at $\Delta$ = -3.53 meV (high photonic content) and $\Delta$ = +0.82 meV (high excitonic content). Red discontinuous lines corresponds to fits based on a single exponential decay ($\Delta$ = -3.53 meV) and on a bi-exponential decay ($\Delta$ = +0.82 meV), respectively. c) Experimentally measured LP lifetime versus $\Delta$ (black dots) and best fit to a two-coupled-oscillator model when including polariton losses by elastic scattering (red line). The main mechanism of LP losses is governed by elastic scattering with negative trions $X^{-1}$, at a rate of $\simeq$ 0.1 scattering events per picosecond at $E_{LP}$ = $E_{X^{-1} }$ = 1.4787 meV, which corresponds to $E_B$ = -1.1 meV. Inset: Photoluminescence spectra of the cavity at $\Delta$ $\ll$ 0, when exciting resonant to acceptor GaAs impurities (1.503 eV). At -1.1 meV from the neutral exciton ($X^0$) a second excitonic transition is visible which we attribute to the negative trion ($X^{-1}$). The tail of the lower energy band corresponds to the cavity mode. Black and red continuous lines corresponds to Lorentzian line fits.}
\label{fig_2} 
\end{center}
\end{figure*}

In order to maximize our chances to observe this phenomenon, we use a semi-integrated, tunable fiber-cavity microscope described previously \cite{Sanchez2013}. It provides a small mode volume comparable or even smaller than other confinement methods, with a nominal Gaussian beam waist of 1.3 $\mu$m, a large finesse (designed to be around 10'000), and efficient optical excitation and collection channels. Figure \ref{fig_1}.c summarizes the main features of the present InGaAs Quantum Well (QW) microcavity.

One of the major experimental challenges is the short lifetime of the polaritons, that mandates the use of ultrafast single photon detectors. In our sample, we measure cavity lifetimes on the order 20 to 40 picoseconds (ps), depending on the cavity-exciton detuning. Given that we expect only a small antibunching signal in the few percent range, even the Avalanche Photodiodes (APDs) with the best time-resolution of around 40~ps will not be good enough to record a clear antibunching signal. While streak cameras can be used to record autocorrelation traces, their extremely low quantum efficiency are a major setback and would lead to extremely long integration times. In order to avoid these issues, we therefore resort to pulsed excitation and slow detectors - a method that was successfully used in the past to demonstrate photon blockade for a quantum dot in a photonic crystal cavity with similarly short decay times \cite{Reinhard2012}. Figure \ref{fig_1}.d summarizes the experimental set-up. 

\begin{figure*}[ht]
\begin{center}
\includegraphics[scale=0.75]{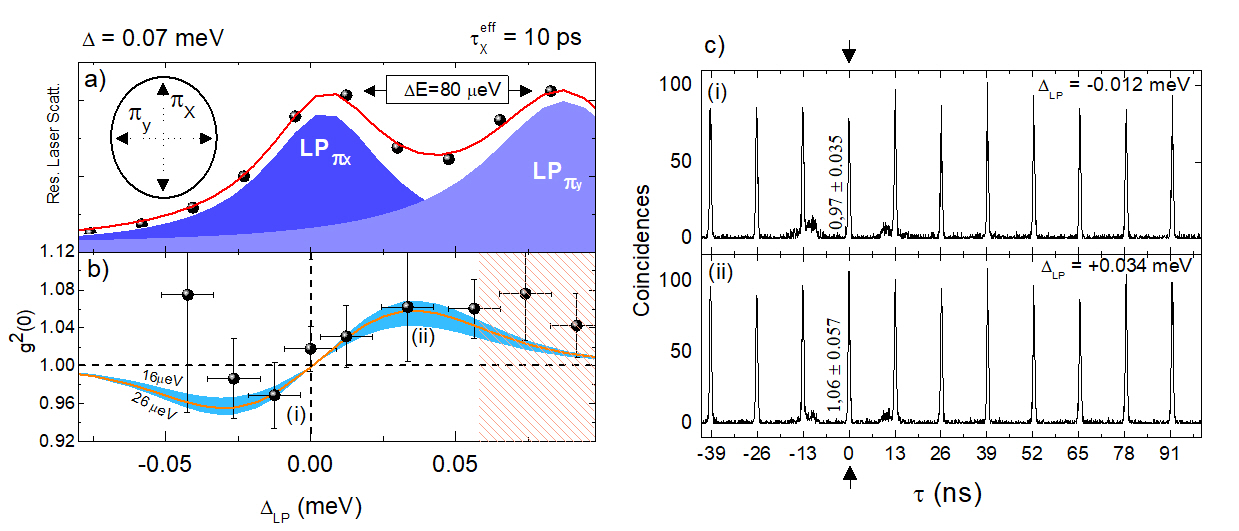} 
\caption{\textbf{Photon autocorrelation data as a function of laser detuning}. a) Resonant transmission data (black dots) on the LP resonance. The LP resonance is split into two linearly polarized cavity resonances, $\pi_{x}$ and $\pi_{y}$. The red continuous line corresponds to the result from a Lorentzian double-peak fit. b) Second-order autocorrelation function at zero time-delay ($g^{(2)}(0)$) as a function of laser detuning $\Delta_{LP}$ from the position of the LP $\pi_{x}$-resonance. The light blue shaded area corresponds to the range of simulated $g^{(2)}(0)$ values for realistic nonlinearities of 16 and 26 $\mu$eV. Orange continuous line corresponds to the best fit of the model, with a nonlinearity of 22 $\mu$eV. The area shaded in brown indicates significant contribution from the $\pi_{y}$ cavity where we expect to see deviations of the measured values from our simulations. c) Raw coincidence histogram from points i) and ii) in b), representing the two different regimes of anti-bunched and bunched photon statistics.}
\label{fig_3}
\end{center}
\end{figure*}

Figure \ref{fig_2}.a displays a low-temperature photoluminescence map of our QW-cavity system as a function of cavity length when pumped off-resonantly at 800 nm. The y-axis is given in units of cavity-exciton detuning ($\Delta =E_C-E_X$), where $E_C$ is the photon cavity energy and $E_X$ is the bare exciton ($X^0$) energy. The avoided crossing between cavity and exciton indicates well-defined polariton states. Fitting with a coupled harmonic oscillators model, we extract a vacuum Rabi splitting between upper (UP) and lower (LP) polariton of $\Omega = 2g \approx 3$ meV. Note that under nonresonant excitation of polariton photoluminescence, a clean avoided crossing in figure \ref{fig_2}.a is only obtained when exciting above the GaAs exciton resonance. In contrast, when exciting between 820-830 nm, the polariton lines appear significantly broadened and we find a second excitonic feature at around $1.1$~meV red-shifted from the bare $X_0$ resonance [see supplementary information]. A similar feature was recently reported in another fiber-cavity setup \cite{Imamoglu2017}. In accordance with literature, we identify the additional line with the negatively charged trion resonance ($X^{-1}$) \cite{Deveaud2005}. This is consistent with the fact that optical excitation of carbon acceptors in the range 820-830 nm provides free electrons \cite{Deveaud2005}, \cite{MunozMatutano2008}. In the context of our experiment, this trion level constitutes a significant energy dependent source of losses and decoherence for polaritons, and thus affects significantly the optimum detuning for which antibunched emission is strongest.

To evaluate quantitatively the underlying loss mechanisms, we performed resonant lifetime measurements as a function of cavity-exciton detuning. Figure \ref{fig_2}.c summarizes the findings. Starting from red-detuning, the lifetime increases when approaching resonance ($\Delta = 0$) as expected from the strong-coupling model and the independently measured lifetimes of cavity and exciton. For slight red-detuning, however, the LP lifetime starts to drastically reduce, indicating a clear deviation from the expected strong-coupling behavior and the presence of additional loss channels \cite{Sermage1996}. For very fast losses, the lifetime curves show a double-exponential decay. Figure \ref{fig_2}.b displays such a bi-exponential decay plus fit at a cavity-exciton detuning of $\Delta = 0.82$~meV, corresponding to an excitonic fraction $|C_X|^2$ = 0.55. The fast decay corresponds to the polariton lifetime of $\tau_{LP} \approx 8$~ps, while we attribute the slow decay to a background from bare excitons with a lifetime of $\tau_X \approx 65$~ps. The likely reason for the fast LP decay is elastic scattering towards localized excitonic states as described in reference \cite{Klembt2017}. The solid line through the lifetime data in Figure \ref{fig_2}.c corresponds to a best fit to the model described in reference \cite{Klembt2017} and in the supplementary information. Output fitting values are cavity lifetime ($\tau_C$ = 16 ps) and peak scattering rate (S = 0.134 ps$^{-1}$). Maximum scattering occurs at $\Delta$ $\approx$ +1 meV. For observing antibunched photon statistics, $\Delta$ has to be chosen such that polariton losses are kept as small as possible, while at the same time polariton-polariton interactions are maximized.  

\begin{figure*}[ht]
\begin{center}
\includegraphics[scale=0.80]{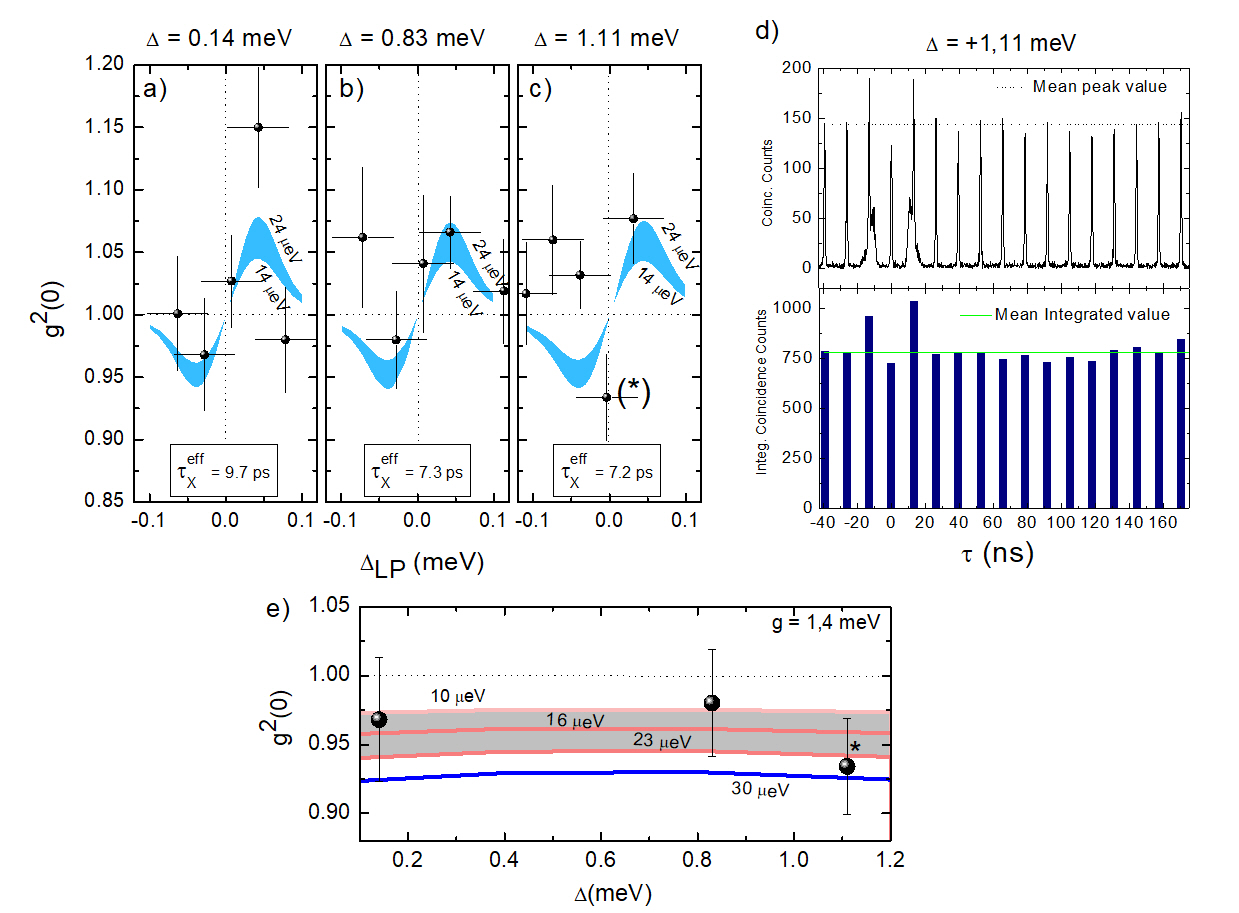} 
\centering
\caption{\textbf{Dependence of photon autocorrelation data on cavity-exciton detuning}. Second-order autocorrelation function at zero time delay, $g^{(2)}(0)$, as a function of laser detuning ($\Delta_{LP}$) for three different cavity-exciton detunings: a) $\Delta$ = 0.14 meV, b)  $\Delta$ = 0.83 meV $\&$ c) $\Delta$ = 1.11 meV. The fiber-cavity configuration used for these measurements has a slightly lower vacuum Rabi splitting ($2g = 2.8$ meV). Black dots are the measured data while the blue region indicates the expected trend extracted from simulations with non-linear constants of 14 and 24 $\mu$eV. d) Upper panel: Raw coincidence counts corresponding to the data point (*) in c); lower panel: total coincidence counts per peak. The black dotted line in the upper panel represents the mean peak value and the green continuous line in the lower panel the mean total count rate per peak. Note that the two pulses at $\pm$ 13~ns were not taken into account in the analysis. e) The black dots correspond to the measured minimum $g^{(2)}(0)$ values from a), b) $\&$ c), while the continuous lines are the model predictions for a range of non-linear constants from 10 - 30 $\mu$eV.}
\label{fig_4}
\end{center}
\end{figure*}

Following this observation and reference \cite{Verger2006}, we chose a cavity-exciton detuning around resonance ($\Delta$ = +0.07 meV) and mapped out $g^{(2)}(0)$ as a function of laser detuning. In a very first step, we performed a resonant transmission measurement for mapping out the position of the LP resonance. Figure \ref{fig_3}.a displays the result. Instead of a single transmission peak we observe a double-peak structure associated with a polarization mode-splitting of around $\simeq$ 80 $\mu$eV into linear $\pi_X$ and $\pi_Y$ modes, an effect well-known from both fiber cavities \cite{Besga2015} and micro-pillars \cite{Reitzenstein2007}. The linewidth of both LP modes is around $\simeq$ 60 $\mu$eV. Based on the resonant linewidth and the corresponding LP lifetime, we then adapted the width of our resonant laser pulses $\delta_L$ accordingly. Selecting the appropriate $\delta_L$ will affect the contrast of the antibunching dip in pulsed excitation correlations. We tested different pulse widths in our theoretical model to find the best strategy for maximizing the contrast in $g^{(2)}(0)$. Details are given in the supplementary information section. 

Using a pulse width of $\delta_L$ = 40 ps and a mean laser power of 12 pW at 76.3 MHz (corresponding to around 0.7 photons/pulse), we recorded $g^{(2)}(0)$ for various laser detunings ($\Delta_{LP}$) across the LP $\pi_X$ mode. The black dots in figure \ref{fig_3}.b represent the data points. They are plotted on top of the theoretical predictions given by our model covering a narrow range of realistic single-mode interaction energies $\hbar\omega_{NL}$ of 16 $\mu$eV to 26 $\mu$eV. This band is shown as the bright blue shaded region in figure \ref{fig_3}.b. The theoretical model is described in the Supplementary Information, and involves a master equation approach taking into account our experimental parameters, the pulsed nature of the excitation, the experimental integration over all possible correlation times, as well as the dark count rates of the detectors ($\sim 80$~cts/s). Exciton losses arising from trion scattering are included in the model by defining an effective excitonic lifetime ($\tau^{\text{eff}}_{X}$) as {\it ad hoc} parameter, which varies with $\Delta$ [see supplementary information]. To determine the most likely $\hbar\omega_{NL}$ of the experimental system, we carried out least squares fitting of the experimental data points in figure \ref{fig_3}.b, with the nonlinearity as a free parameter. For the fitting procedure, the two points in the brown shaded area were omitted as they contained large contributions from the second ($\pi_{y}$) polariton mode. The best fit to the theoretical model was obtained with a nonlinearity of $\hbar\omega_{NL}$ = 22 $\mu$eV (orange continuous line in figure \ref{fig_3}.b), where the coefficient of determination of the fitting is $R^2=0.72$ (see Methods). 

As predicted by the simulations, the experimentally determined $g^{(2)}(0)$ clearly varies when changing from red to blue laser detuning. For small laser red-detuning ($\Delta_{LP}$ = - 0.012 meV), we find a value of $g^{(2)}(0)= 0.97 \pm 0.04$ which is slightly below the shot noise ($g^{(2)}(0)= 1$), indicating the underlying quantum nature of the LP quasiparticles. $g^{(2)}(0)$ then rises quickly above 1 as the laser pulse is detuned to the blue side of the LP resonance. At $\Delta_{LP}$ = + 0.034 meV, it reaches a maximal value of $1.06 \pm 0.06$, consistent with our model. 

Figure \ref{fig_3}.c displays the raw correlation data corresponding to points (i) and (ii) of the plot. The zero delay peak characterizes the ultrafast quantum correlations we are looking for, while the other peaks originate from large-delay uncorrelated photon coincidences. The total coincidence counts in each peak were evaluated by counting coincidences within the Gaussian width of the pulses, set by the time resolution of the detection system. The experimental uncertainty in $g^{(2)}$(0) was determined from the standard deviation in the total counts of all the uncorrelated peaks. Note that unwanted APD cross-talk~\cite{Reinhard2012},\cite{Wood2018} contributes to the two peaks at $\pm$ 13 ns. We thus discarded them in the ensuing statistical analysis. 

Upon blueshifting the laser by more than half a linewidth with respect to the LP $\pi_x$ mode, its high energy tail starts to overlap significantly with the upper lying LP $\pi_y$ mode. While both modes are cross-polarized, we do not have precise control over the polarization state of the excitation to make sure that both mode are never excited simultaneously. As a result, if we start pumping the upper energy mode, even negligibly, it might add uncorrelated photons to the signal. We thus expect the blockade mechanism to fail in this case (light-brown shaded area in figure \ref{fig_3}.b).

In a second series of measurements we explored different cavity-exciton detunings ($\Delta$) for maximizing the achievable antibunching,  measuring $g^{(2)}(0)$ versus $\Delta$. Figure \ref{fig_4} summarizes the results. Due to the very long integration time required, we measured only five different laser detunings ($\Delta_{LP}$) for each of the three cavity-exciton detunings $\Delta = 0.14$, $0.83$ and $1.11$~meV. As $\Delta$ increases, polariton scattering losses broaden the LP line (60 $\to$ 140 $\mu$eV, see supplementary information), reducing the number of emitted photons and the effective excitonic lifetime ($\tau^{eff}_{X}$). To obtain a minimally manageable count rate of 1000 counts/s per APD, we increased the resonant excitation power accordingly (up to 40 (160) pW for $\Delta$ = 0.14 $\&$ 0.83 meV (1.11 meV)). This higher excitation power compensates for the increased polariton losses, such that the intracavity effective population remains around 1. Indeed, considering the scattering rate $\gamma_{L}(\Delta)$ analyzed in figure \ref{fig_2}.c, with values between 0.1 $-$ 0.13 polaritons/ps and a laser pulse width of $\delta_L$ $\simeq$ 40 ps, polariton losses could remove $\simeq$ 4 $-$ 5 polaritons/pulse when the cavity is tuned to $\Delta$ $\simeq$ 0.15 to 1 meV. 

Figure \ref{fig_4}.a$-$c, displays the measured (black dots) and simulated (blue shaded area) $g^{(2)}(0)$ for the three different cavity-exciton detunings. Simulations and measured values agree best within a range of non-linear constants of $\hbar\omega_{NL}$ = 14-24 $\mu$eV. Due to technical reasons, the experimental uncertainty in laser detuning was $\sim$ 40 $\mu$eV for this set of measurements, slightly larger than before ($\sim$ 10 $\mu$eV in figure \ref{fig_3}). In spite of this, the data qualitatively follow the expected dispersive shape of $g^{(2)}$(0)$(\Delta_{LP})$. We note that with increasing $\Delta$, the experimental $g^{(2)}$(0) values exhibit a small positive offset, which might be due to the increased polariton loss rate, as it has been explained above. Figure \ref{fig_4}.d displays the raw coincidences plot corresponding to the point labelled (*) in Fig.\ref{fig_4}.c, observed at $\Delta$ = +1.11 meV and $\Delta_{LP}$ $\simeq$ 0. The black dotted line shows the mean value of uncorrelated coincidences (calculated without the zero-delay peak and the $\pm$ 13 ns peaks). The zero delay peak lies clearly below this line. This trend is confirmed by the quantitative analysis shown in the bottom panel of figure \ref{fig_4}.d. We find a value of $g^{(2)}$(0) = 0.934 $\pm$ 0.035, corresponding to 1.88 $\sigma$ away from $g^{(2)}$(0) = 1 (94$\%$ of confidence). Finally, figure \ref{fig_4}.e displays the minimum measured $g^{(2)}$(0) value (dots) for each of the three set of data in a$-$c. For comparison, we underlay simulated trends (continuous lines) for different values of the non-linear constant ($\hbar\omega_{NL}$= 10; 16; 23; 30 $\mu$eV). Experiment and theory seem to best match for $\hbar\omega_{NL}$ = 10 $-$ 23 $\mu$eV.

Next, we compare the measured nonlinearity with other reports in literature. The single mode non-linear interaction constant can be expressed as $\hbar\omega_{NL}$=$\hbar\kappa\int dxdy|\varphi_C(x,y)|^4$ = $\hbar\kappa/A$, where $\hbar\kappa$ represents the exciton-exciton interaction constant and $A=1/\int dx|\varphi_C(x)|^4$ denotes the effective polariton mode area set by the cavity \cite{Verger2006}. $\hbar\kappa$ can be calculated from the zero-wavevector value of the Coulomb exchange interaction \cite{Ciuti2003, Ciuti1998}. For our InGaAs QW with 4$\%$ indium, we expect $\hbar\kappa\simeq$ 14.1 $\mu$eV $\times$ $\mu m^2$ \cite{Verger2006, Ciuti1998}. This is in line with recent reports based on measuring sound velocity in a polariton fluid \cite{Amo2009} and on studying non-resonantly pumped polariton condensates in micropillars \cite{Ferrier2011}. Assuming a Gaussian mode profile of our fiber cavity with an estimated waist of $\omega_0 = 1.3$ $\mu$m, we find a lower theoretical value for the single mode nonlinearity of $\hbar\omega_{NL}$ $\simeq$ 3 $\mu$eV. Among several possible explanation for this discrepancy, we suspect in particular that the mode waist $\omega_0$, which contributes quadratically to the nonlinearity, is overestimated \cite{Dufferwiel2014}. A proper calculation of $\omega_0$ should indeed account for the strong coupling regime that increases significantly the mode effective index\cite{Richard2015}.

Finally, we comment on the possibility of reaching the polariton blockade regime for which the system nonlinearity $ U = \frac{\hbar\omega_{NL}}{4}$ \cite{Yamamoto1999} exceeds the polariton linewidth $\hbar\gamma_{LP}$, i.e. $U > \hbar\gamma_{LP}$ \cite{Verger2006}. Defining $\eta_{BKD} = \frac{U}{\hbar\gamma_{LP}}$, our experiments returns $\eta_{BKD}$ $\simeq$ 0.07 $-$ 0.10 (at $\Delta$ = +0.07) and $\eta_{BKD}$ $\simeq$ 0.03 $-$ 0.04 (at $\Delta$ = + 0.82), which is more than one order of magnitude below $\eta_{BKD} = 1$, and hence deep in the {\it weak blockade} regime. Note that these values are in good agreement with the non-linearity range found in a similar open fiber cavity system \cite{Imamoglu2017}. In order to reach the strong blockade regime, a factor $\approx$ 4 times smaller mode waist is needed. This would require either a combination of our fiber-cavity system with nanophotonic elements such as solid-immersion lenses or the use of fully integrated nanophotonic structures, such as photonic crystal cavities. Alternatively, resonant enhancement of interactions via the polaritonic Feshbach resonance effect \cite{Carusotto2010, Deveaud2014} or the unconventional photon blockade effect \cite{Flayac2017} might be feasible.

In summary, several ingredients made the observation of antibunched light from confined quantum-well cavity polaritons possible: the tight optical confinement and the high-quality mirrors of the fiber cavity microscope, the understanding (and avoidance) of the additional loss channels due to trion and QW exciton scattering, the experimental technique of pulsed excitation with lifetime-tailored laser pulses and finally the extremely low excitation powers leading to tens of hours of integration time per data point. In order to make practical use of the quantum nature of polaritons, however, further engineering of polariton non-linearities is required. As an alternative, other semiconductor platforms, such as 2D single layered materials with their strong excitonic effects \cite{Low2017,Dufferwiel2017} might ultimately provide a viable path forward. With strong polariton quantum non-linearities at hand, polariton-based high repetition rate single-photon emitters, Fock state optical switchers, all-optical quantum gates, and light-based 2D driven dissipative quantum simulators would be possible. 

\subsection*{Methods}

\subsubsection{Fiber cavity microscope} 
The home-built fiber cavity setup is sitting in a liquid-helium dewar at 4K. The cavity itself consists of a curved fiber-based top mirror with a highly-reflective dielectric Bragg coating (Laseroptik GmbH, Germany) with a designed finesse of a few 10'000. Its small curvature radius (R = 13 $\mu$m) ensures tight mode confinement. The depth of the curved mirror is $\rho$ = 1.3 $\mu$m. The second cavity mirror is an MBE-grown AlAs/GaAs DBR mirror with a single In$_{0.04}$Ga$_{0.96}$As quantum well (QW) on top. Sample and fiber tip are mounted on attocube positioners for controlling the lateral position and distance of the fiber tip relative to the sample. For all photon correlation measurements described here, the effective cavity length was calculated to be around $4.8$ $\mu$m, taking into account the effective penetration depth into the AlAs/GaAs mirror \cite{Besga2015} and the depth of the curved mirror ($\rho$). Assuming a Gaussian mode shape, we calculate an effective cavity mode waist ($\omega_0$) at the position of the InGaAs QW of approximately $1.3$~$\mu$m. 

\subsubsection{Resonant transmission measurements}
For continuous-wave resonant transmission measurements, we use a highly stable wavelength-tunable cw Ti:Sapphire laser (SolsTiS from M-Squared). The power sent into the fiber cavity is monitored carefully. The light transmitted through the fiber cavity is directed onto silicon APDs for detection (SPCM-AQRH-14 from Excelitas). Photons are counted using the correlator box PicoHarp 300 from PicoQuant. For time-resolved and pulsed resonant correlation measurements a pulsed Ti:Sapphire laser (MIRA) with 76.3 MHz repetition rate and 4 ps pulsewidth is used for excitation. Resonant lifetime traces are recorded on a streak camera with a temporal resolution of around 2 ps (OptoScope from Optronis GmbH). Conventional photoluminescence is excited with the cw Ti:Sapphire laser, and detected on the silicon CCD (Pixis 100) attached to the monochromator with 1500 gr/mm grating (Acton Spectra Pro 2750). Photon autocorrelation data are measured with a conventional free-space Hanbury-Brown and Twiss interferometer. The two free-space silicon APDs have a quantum efficiency of $\sim$ 40$\%$ at 830 nm and a time resolution of around $\sim$ 350 ps.   

\subsubsection{Pulse shaping}
In order to adapt the resonant probe pulses to the measured resonant polariton lifetimes, we shape the temporal width of the intrinsic laser pulses from the MIRA laser by sending them through the monochromator. The effective numerical aperture is controlled via an iris before the entrance slit of the monochromator and determines the effective laser pulse width at the monochromator output. The pulses are then coupled into a single-mode optical fiber for delivery to the fiber cavity. The single-mode fiber at the monochromator output acts as a pinhole. The monochromator grating angle controls the centre wavelength of the laser pulse. The method provides effective tuning of the laser pulse widths between 15 and 50 ps [see supplementary information]. The pulse shapes are monitored on the streak camera, the pulse centre wavelengths are read-out using a wavemeter from High Finesse (WSU-10). 

\subsubsection{Agreement between data and model}
In order to have a quantitative estimation for the deviation between the model and the data, we define the so-called coefficient of determination as:
\begin{equation*}
R^2 = 1 - \frac{\sum(g^2_{i}(0)_{Exp}-g^2_{i}(0)_{Mod})^2}{\sum(g^2_{i}(0)_{Exp}-1)^2} 
\end{equation*}
Where $g^2_{i}(0)_{Exp}$ denotes the measured values for $g^2(0)$, $g^2_{i}(0)_{Mod}$ are the numerical predictions for $g^2(0)$ and the '1' in the denominator represents $g^2(0) = 1$ for uncorrelated photons. $R$ directly reflects a measure for how much the data are consistent with strongly interacting polaritons versus completely uncorrelated polaritons. $R=1$ indicates 100$\%$ match between data and model.

\subsection*{Acknowledgements}
We thank Scott Martin and Dirk Taylor from the Commonwealth Scientific and Industrial Research Organisation (CSIRO) (Lindfield - New South Wales) for their technical support. This work was funded by the Australian Research Council Centre of Excellence for Engineered Quantum Systems (CE110001013).

\subsection*{Author contributions}
\textbf{G.M.M.} and \textbf{A.W.} carried out the spectroscopy and photon correlation experiments and analyzed the data. \textbf{M.J.}, with the help of \textbf{B.Ba.}, implemented the master equation model and contributed to the analysis of the data. \textbf{G.M.M.}, \textbf{A.W.} and \textbf{M.J.} have equally contributed to this research. \textbf{X.V.A.} built the spectroscopy set-up. \textbf{A.R.} and \textbf{B.Be.} built the cavity microscope and designed the laser machining system for making the fiber cavities. \textbf{A.L.}, \textbf{J.B.} and \textbf{A.A.} provided the quantum well sample and discussed the underlying polariton physics. \textbf{M.R.} and \textbf{T.V.} conceived the central idea of the work and related experiments, supervised the experimental work and contributed to discussions. The manuscript was written by \textbf{G.M.M.} and \textbf{T.V.} with varying contributions from all other authors.

\subsection*{Competing Interests}
The authors declare that they have no competing financial interests.
\subsection*{Correspondence}
Correspondence and requests for materials should be addressed to G.M.M. ~(email: guillermo.munozmatutano@mq.edu.au) and T.V. ~(email: thomas.volz@mq.edu.au).

\setcounter{figure}{0}
\renewcommand\thefigure{S.\arabic{figure}}   

\section*{Supplementary Information}

\subsection{Theoretical model}

The starting point for our model is a quantum well exciton coupled to a planar microcavity photon mode. The non-dissipative dynamics of such a system are governed by a Hamiltonian given by \cite{Verger2006}:
\begin{align}
H = \int d{\bf{x}} \sum_{i,j \in \{ X,C \}} \hat{\Psi}_i^{\dagger}({\bf{x}}) h^0_{i,j} \hat{\Psi}+ \nonumber \\ +\frac{\hbar \kappa } {2} \, \int d {\bf{x}} \hat{\Psi}_X^{\dagger}({\bf{x}}) \hat{\Psi}_X^{\dagger}({\bf{x}}) \hat{\Psi}_X({\bf{x}}) \hat{\Psi}_X({\bf{x}}) +\nonumber \\
     + \frac{\hbar \Omega_R } {n_{\text{sat}}} \, \int d {\bf{x}} \hat{\Psi}_C^{\dagger}({\bf{x}}) \hat{\Psi}_X^{\dagger}({\bf{x}}) \hat{\Psi}_X({\bf{x}}) \hat{\Psi}_X({\bf{x}}) + H.c. +\nonumber \\
     + \hbar \int d {\bf{x}} \, F({\bf{x}}, t) e^{-i \omega_p t} \hat{\Psi}_C^{\dagger}({\bf{x}}) + H.c.
\label{eqFullH}
\end{align}
Here $\hat{\Psi}_{C,X}$ are the spatially-dependent quantum field annihilation operators that describe excitons ($X$) and cavity photons ($C$) and the kinetic energy term of the planar microcavity is given by
\begin{equation}
h^{0} = \hbar \left( \begin{array}{cc}
\omega_X(-i \nabla) & \Omega_R \\
\Omega_R & \omega_C(-i \nabla) + V_C({\bf{x}}) 
\end{array} \right)
\end{equation}
where $V_C({\bf{x}})$ is the photon confining potential and $\hbar \omega_{C}$ and $\hbar \omega_{X}$ are the photon and exciton energies respectively. $\Omega_R$ is the vacuum Rabi frequency, giving the coupling rate between photons and excitons. There are two nonlinear exciton-exciton interactions: The first is a pure exciton-exciton coupling given by $\kappa$, and the second is the anharmonic photon-exciton coupling governed by the exciton oscillator strength saturation density $n_{\text{sat}}$. The photonic pumping term is given by $F_p$ and is assumed to be monochromatic with a frequency $\omega_p$.

In the case of strong photonic confinement, the only relevant modes are the fundamental cavity photon mode with spatial wavefunction $\phi_c({\bf{x}})$ and the exciton mode with the same spatial shape \cite{Verger2006}. This simplifies the Hamiltonian (\ref{eqFullH}) to

\begin{align}
H = & \,\, \hbar \omega_X b^{\dagger} b + \hbar \omega_C a^{\dagger} a + \hbar \Omega_R ( b^{\dagger} a + b a^{\dagger}) + \nonumber \\ & + \frac{\hbar \omega_{\text{nl}}}{2} b^{\dagger} b^{\dagger} b b - \hbar \alpha_{\text{sat}} \Omega_R (b^{\dagger} b^{\dagger} a + a^{\dagger} b b b) +\nonumber \\
    & + \hbar {\cal{F}}(t) e^{-i \omega_p t} a^{\dagger} + \hbar {\cal{F}}^{*}(t) e^{i \omega_p t} a
\end{align}
where $a$ and $b$ are bosonic annihilation operators of the cavity photon mode and the exciton mode respectively. The effective photon drive strength ${\cal{F}}$, and effective nonlinearities $\alpha_{\text{sat}}$ and $\omega_{\text{nl}}$ are given by ${\cal{F}}(t) = \int d {\bf{x}} F_p({\bf{x}},t) \phi_C^*({\bf{x}})$, $\alpha_{\text{sat}} = \int d {\bf{x}} | \phi_C({\bf{x}}) |^4/n_{\text{sat}}$, and $\omega_{\text{nl}} = \kappa \int d {\bf{x}} | \phi_C({\bf{x}}) |^4 $.

Finally, we note that the saturation parameter is negligible for typical microcavity parameters\cite{Ciuti2003} and go into the interaction picture to remove the bare photon and exciton energies. This yields the effective Hamiltonian
\begin{align}
H_{\text{eff}} = & \hbar \, \Delta \omega_X b^{\dagger} b + \hbar \, \Delta \omega_C a^{\dagger} a + \hbar \Omega_R (b^{\dagger} a + a^{\dagger} b) + \nonumber \\ 
+ & \frac{\hbar \omega_{\text{nl}}}{2} b^{\dagger} b^{\dagger} b b + \hbar {\cal{F}}(t) (a^{\dagger} + a)
\label{eqTwoModeHamiltonian}
\end{align}
where without loss of generality we have taken ${\cal{F}}(t)$ to be real and defined the detunings $\Delta \omega_c = \omega_C - \omega_p$ and $\Delta \omega_X = \omega_X - \omega_p$. In order to make contact with the experimental parameters $\Delta$ and $\Delta_{LP}$ used in the main text, we note that $\Delta \omega_C = \Delta/2 - \Delta_{LP} + \sqrt{\Omega_R^2 + \Delta^2/4}$ and $\Delta \omega_X = \Delta \omega_C - \Delta$.

The Hamiltonian (\ref{eqTwoModeHamiltonian}) describes a system consisting of a single-mode photon field ($a$) and a single-mode exciton field ($b$). The two fields are coupled with a Rabi frequency $\Omega_R$, and the exciton field has a self-coupling nonlinearity with strength $\omega_{\text{nl}}$. The photon field has a time-dependent drive ${\cal{F}}(t)$ which is related to the laser power via ${\cal{F}}(t) =  \sqrt{P(t)\gamma_C / \hbar \omega_p}$ where $P(t)$ is the input power into the cavity mode and $\gamma_C$ is the optical cavity decay rate.

In order to study the cavity dynamics and photon correlation functions it is also necessary to include dissipation and loss. This requires a master equation description of the system. We denote the density matrix of the system by $\rho(t)$, and introduce losses $\gamma_C$ and $\gamma_X$, corresponding to the homogeneous linewidths of the photons and excitons respectively, resulting in a master equation for the system given by 
\begin{align}
\frac{d \rho}{dt} = & \frac{i}{\hbar} \left[  \rho, H_{\text{eff}} \right] + \frac{\gamma_C}{2} \left( 2 a \rho a^{\dagger} - a^{\dagger} a \rho - \rho a^{\dagger} a \right) + \nonumber \\ 
+ & \frac{\gamma_X}{2} \left( 2 b \rho b^{\dagger} - b^{\dagger} b \rho - \rho b^{\dagger} b \right).
\label{eqMasterEquation}
\end{align}
Due to the fact that the photonic driving term describes a Gaussian pulse rather than continuous excitation, we must find the full time-dependent solution to (\ref{eqMasterEquation}) and not merely find the steady state solution. To do this we treat the master equation as a time-dependent matrix differential equation, and numerically solve it in a tensor product Fock basis for the operators $a$ and $b$. We truncate the basis by limiting the number of excitations in both the photon and exciton modes. For parameters describing our experiment, we found that an acceptable cutoff was between five and eight excitations in each mode depending on laser power.
\begin{figure*}
\begin{center}
\includegraphics[width=12cm]{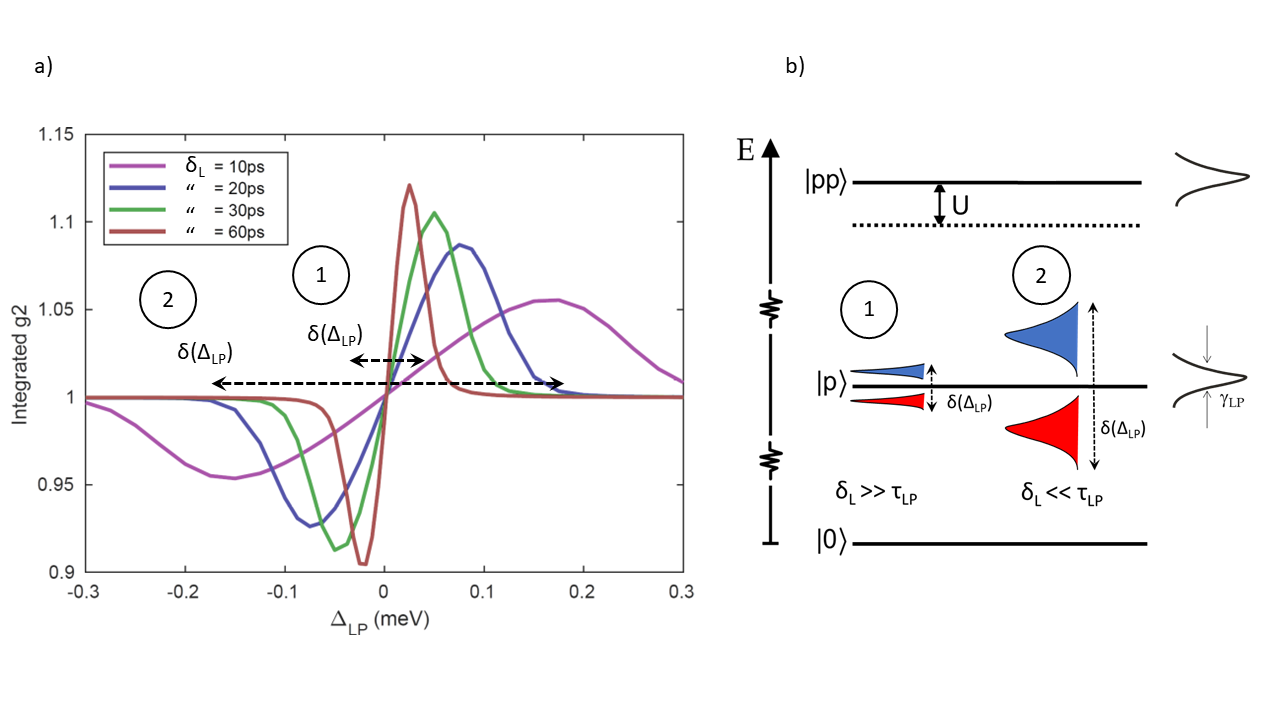}
\centering
\caption{\textbf{$g^{(2)}(0)$ as a function of laser detuning}. a) Theory model output using different pulsed laser widths to probe the polariton blockade effect. b) Energy diagram for two pulsed laser width scenarios. In both figures two different excitation conditions are labeled: 1 - the pulsed laser width is longer than the lower polariton lifetime; and 2 - where the laser pulse width is shorter than or similar to the lower polariton lifetime.}
\label{fig_S1}
\end{center}
\end{figure*}

Given the time-dependent solution $\rho(t)$, it is possible to use the quantum regression theorem to find two-time correlation values. For example, the two-time second-order coherence function for the photon field is given by
\begin{align}
G^{(2)}(t,t') = & \text{Tr} \left( a \, U(t,t') \left[ a \, \rho(t') a^{\dagger} \right] a^{\dagger} \right)
\end{align} 
and the normalized version by 
\begin{align}
g^{(2)}(t,t') = & \frac{\text{Tr} \left( a \, U(t,t') \left[ a \, \rho(t') a^{\dagger} \right] a^{\dagger} \right) } {\text{Tr} \left( a \, \rho(t) a^{\dagger} \right) \text{Tr} \left( a \, \rho(t') a^{\dagger} \right) }
\end{align}
where $U(t,t')$ is the evolution superoperator that acts on the density matrix as $\rho(t) = U(t,t') \rho(t')$.

However, we note that the experiment does not find the actual function $g^{(2)}(t,t')$ within the laser pulse, as the pulse width is shorter than the time resolution of the detectors used. Rather, over many pulses, the experiment registers all possible coincidences between $t$ and $t'$ and bins them together. In this situation the correct quantity to compare to experiment is \cite{Reinhard2012}:
\begin{equation}
\bar{g}^{(2)}(0) = \frac{\displaystyle 2 \int^{T}_{-T} dt_1 \int^T_{t_1} dt_2 \,  G^{(2)}(t_1, t_2)} {\displaystyle \int^{T}_{-T} dt_1 \int^{T}_{-T} dt_2 \, I(t_1) I(t_2)}
\label{eqPulseAveraging}
\end{equation}
where the limits $(-T, \, T)$ encompass the full duration of each pulse, and $I(t) = {\text{Tr}} \left[ \rho(t) \, a^{\dagger} a \right] $ is the average photon flux at time $t$.

The quantity $\bar{g}^{(2)}(0)$ corresponds to the quantity that would be measured in the experiment if the detectors were perfect and noiseless. Any real detectors, of course, are not perfect.
\begin{figure}
\begin{center}
\includegraphics[width=7.5cm]{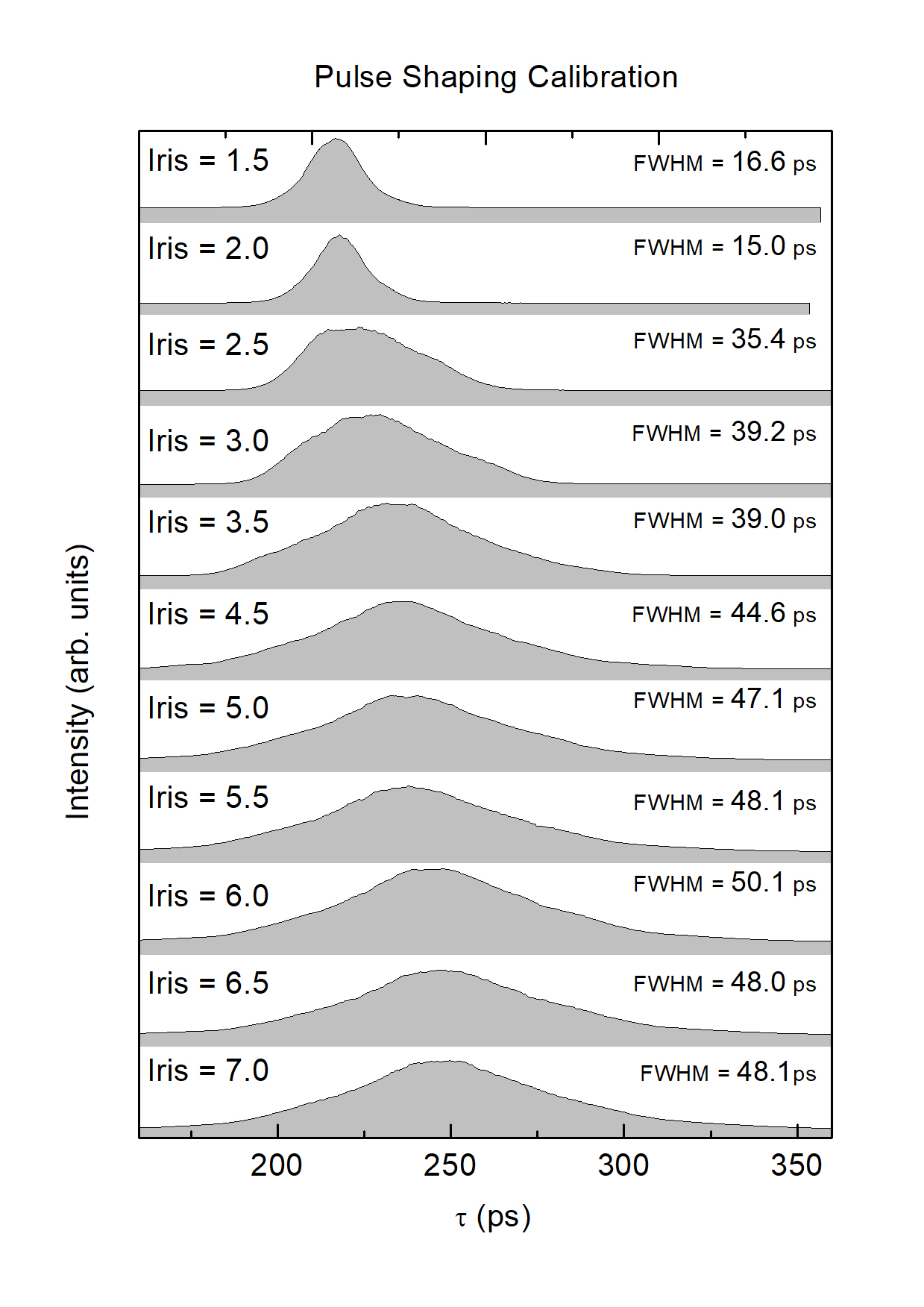}
\centering
\caption{\textbf{Measured laser pulse width}. Streak Camera traces of laser pulses with different values of the incoming monochromator numerical aperture. With this pulse shaping technique it is possible to change the effective pulse width $\delta_L$ from 15 to 50 ps.}
\label{fig_S2}
\end{center}
\end{figure}
In our experiment, the greatest contribution to noise was the detector dark count rate, which was estimated at 80 counts per second. In order to include the dark counts, the signal-to-noise background from these dark counts was measured over the full range of $\Delta_{LP}$ for the detunings $\Delta$ of interest. This background was then modeled as detector events with Poissonian statistics which were added as a separate source in Eq. (\ref{eqPulseAveraging}). It is this final version of $\bar{g}^{(2)}(0)$ that gives the theory values used in the main text.

The simulation parameters we used were $\Omega_R = 3.0$\,meV, $\gamma_C = 41.1\,\mu$eV (16\,ps), and $\gamma_X = 62.1\,\mu$eV (10.6\,ps), with laser powers of 10\,pW -- 100\,pW. We note that over this power range there is almost no change in predicted correlation values. The nonlinearity $\omega_{\text{nl}}$ was left as the single free fitting parameter, with a value of $\omega_{\text{nl}} \approx 22\,\mu$eV providing the best fit. 
 
\subsection{Laser pulse shaping dynamics}

The correlation measurements presented in this paper utilized a picosecond pulsed laser with external pulse manipulation. This pulse manipulation involved spectral shaping of the pulses to increase their duration. The process is illustrated in Figure S.1 and the pulse shaping set up is explained in the main text. With this technique, laser pulses ranging from 15-50 ps widths ($\delta_{L}$) could be achieved.

Longer pulses have two useful effects. First, longer pulses yield a narrower pulse spectrum allowing the selective addressing of points of interest across the lower polariton (LP) linewidth ($\gamma_{LP}$). But, more importantly, our theoretical model shows (Fig S.1.a) that the temporally broadened pulses shift the minimum of the $g^{(2)}(0)$ distribution closer to the LP resonance ($\Delta_{\text{LP}}$ = 0), as well as increasing the degree of antibunching.

The better antibunching is due to the the blockade becoming more effective as the spectral width of the pulse shrinks below the nonlinear energy shift imparted due to the presence of a second polariton. The shift in the minimum is beneficial as it moves the peak $g^{(2)}(0)$ antibunching into the region where there is better signal (i.e. more counts due to being near resonance), enhancing the signal to noise ratio of the measurement. 

This is illustrated in Figure S.1.b, which shows an energy diagram with two different pulsed excitation conditions. "1" corresponds to a situation where $\delta_{L} > \tau_{LP}$, so that the spectral width of the pulse is less than $\gamma_{LP}$. "2" corresponds to the opposite situation, were $\delta_{L} \le \tau_{LP}$, so the pulse spectral width is on the order of or greater than $\gamma_{LP}$. This results in a compressed (1) or expanded (2) excitation laser range ($\delta(\Delta_{LP})$) where the antibunching / bunching photon statistics are observed. 

To manipulate $\delta_L$ the laser light is sent through a grating monochromator with a variable incoming iris diameter, which controls the input numerical aperture. By increasing the iris diameter, the effective numerical aperture of the monochromator expands, illuminating more of the grating. Pulse widths from 15 $\rightarrow$ 50 ps could be readily achieved and accurately measured with a streak camera (Figure S.2). 

\subsection{Negative Trion identification}

In the preliminary work for the results presented above, characterization of the polariton system showed features that indicated the presence of other excitonic states. Photoluminescence (PL) spectra of the avoided crossing under 825 nm pumping show a small but distinct spectral feature approximately 1.1\,meV red detuned from the bare exciton (X$^0$) resonance which can be seen in Fig S.3. A similar feature has been reported before in fiber based cavity experiments \cite{Imamoglu2017} and has been explained from the strong coupling of the cavity mode to a impurity resonance in the semiconductor. However, when compared with literature\cite{Deveaud2005}, this feature agrees strongly with the negative trion state (X$^{-1}$). This argument is supported through the inclusion of losses from scattering off the trion state in the analysis of the resonant lifetimes.
\begin{figure}
\begin{center}
\includegraphics[width=8cm]{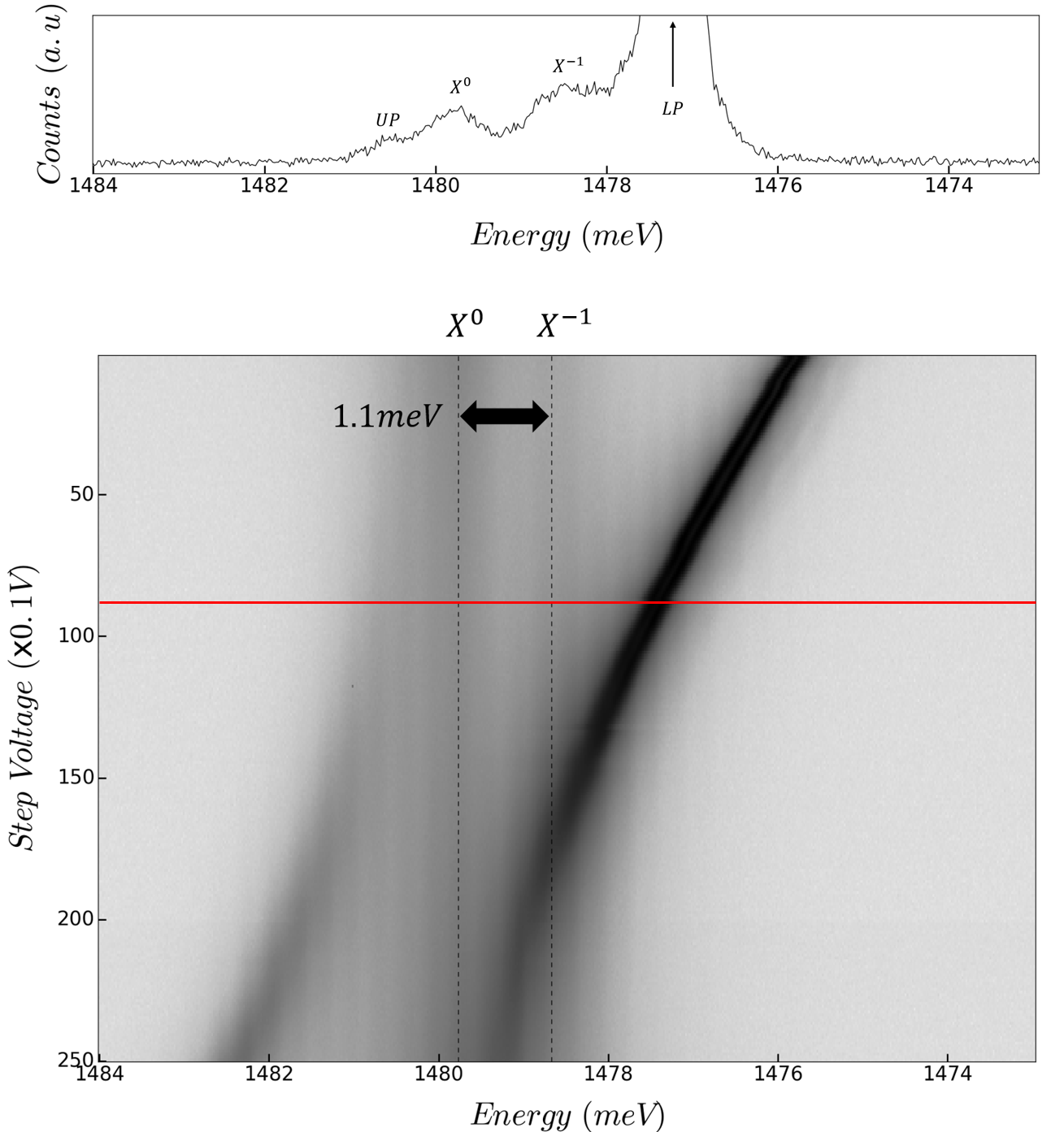}
\centering
\caption{\textbf{Polariton avoided crossing when pumped at 825nm}. The black dotted lines show the neutral exciton ($X^0$) and negative trion ($X^{-1}$) states. The top spectrum is a slice taken from the avoided crossing with its position marked with the red line. The $X^{-1}$ state is approximately 1.1 meV red detuned from the neutral exciton state.}
\label{fig_S3}
\end{center}
\end{figure}
\begin{figure}
\begin{center}
\includegraphics[width=8cm]{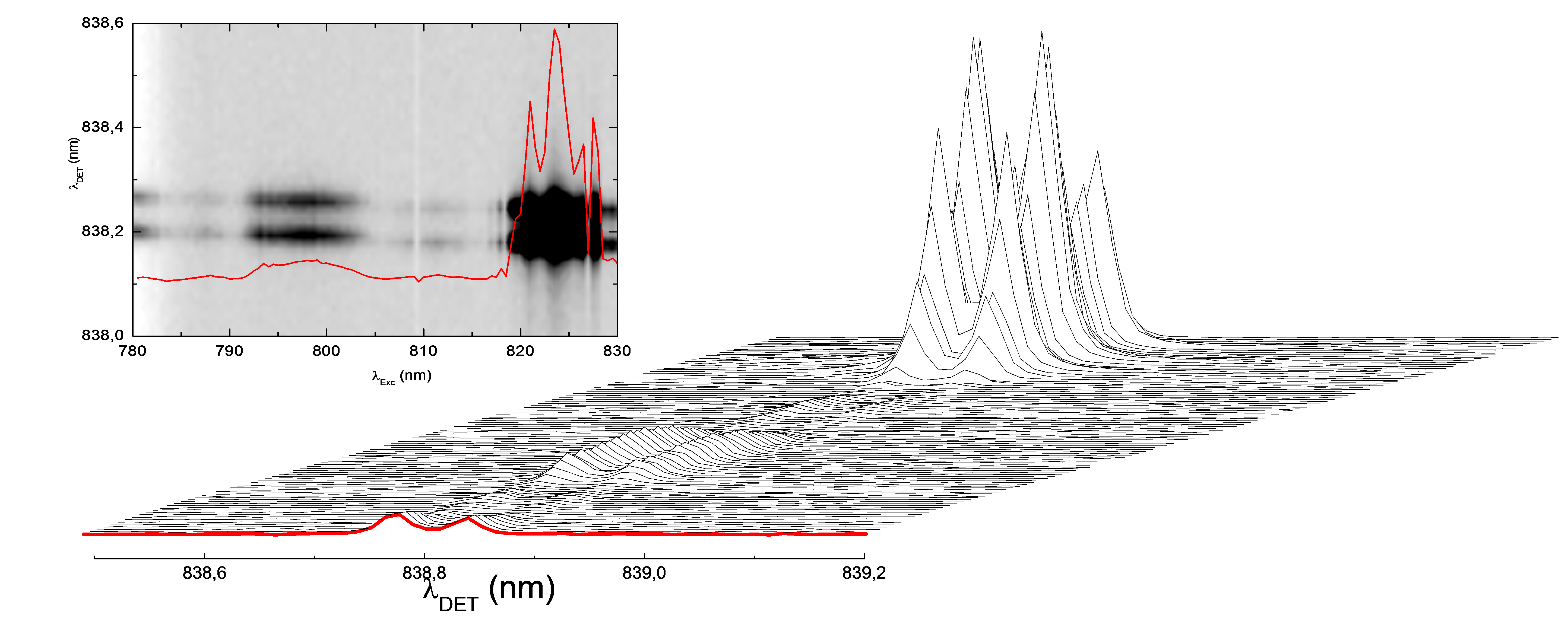}
\centering
\caption{\textbf{Excitation Photoluminescence at fixed cavity-exciton detuning}. The main waterfall graph and inset graph show two distinct pumping ranges for cavity emission. The  800 nm excitation "window" corresponds to optical pumping above the GaAs exciton resonance, while the 825 nm excitation "window" corresponds to optical pumping close to the GaAs carbon acceptor impurity band. Although lower in emitted intensity, the 800 nm excitation window does not excite free electrons from carbon impurities to form trions.}
\label{fig_S4}
\end{center}
\end{figure}

\begin{figure}
\begin{center}
\includegraphics[width=9cm]{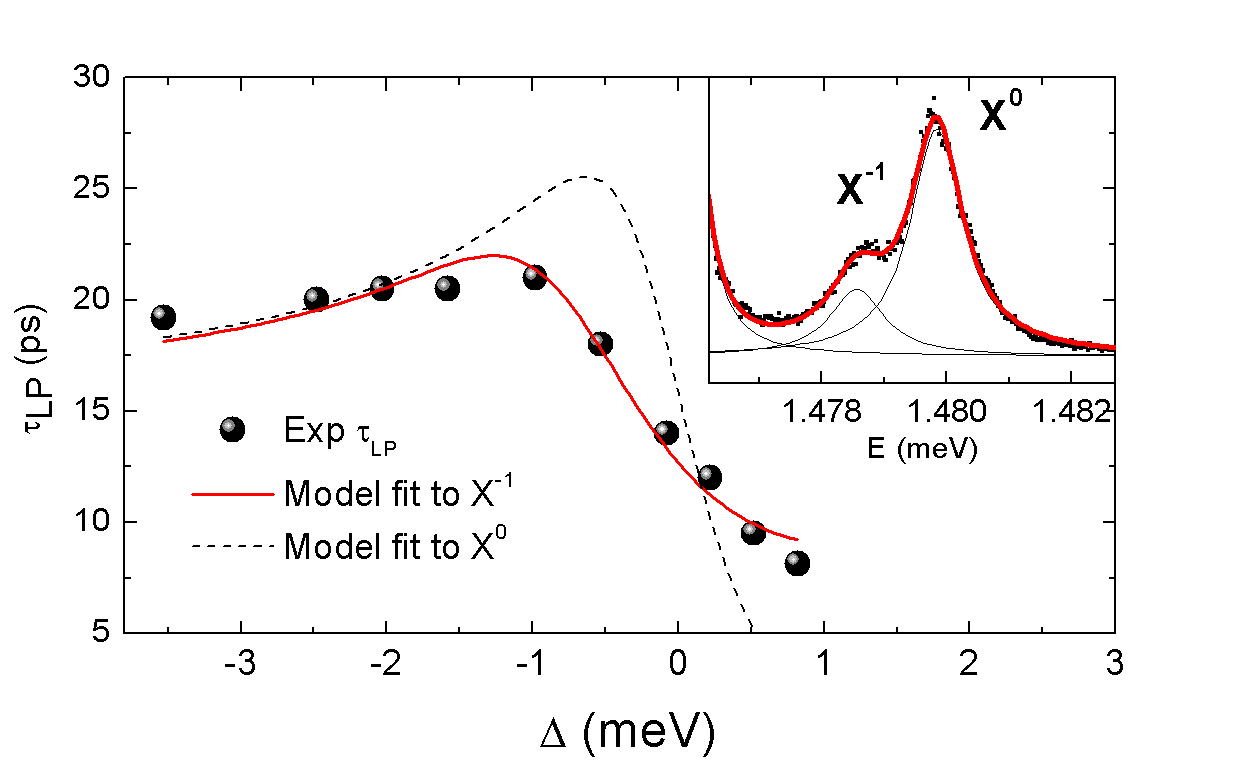}
\centering
\caption{\textbf{Polariton lifetime with scattering losses}. Experimentally measured Lower Polariton (LP) lifetime vs $\Delta$ (black dots) and best fit to a two-coupled-oscillator model when including polariton losses by elastic scattering when using the negative trion $X^{-1} $ (red line) and neutral exciton $X^0$ (black dotted line) as the main source of losses. Inset:  Photoluminescence spectra when exciting resonant to GaAs impurities. Lorentzian fits (black lines) of the neutral exciton and negative trion illustrates their respective contributions.}
\label{fig_S5}
\end{center}
\end{figure}

\begin{figure}
\begin{center}
\includegraphics[width=8cm]{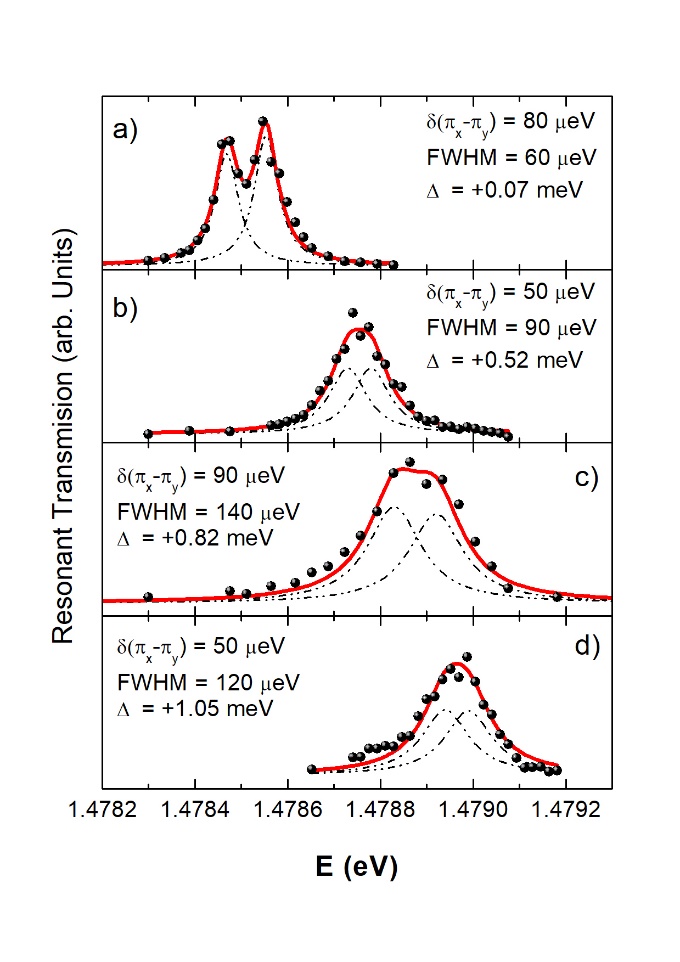}
\centering
\caption{\textbf{Lower Polariton transmission spectra at different exciton detunings.} Polarisation splitting of the lower polariton mode is fit with two Lorentzians (black curves under red curve) with equal area and linewidth, demonstrating equal contributions of each polarisation to the polariton mode. FWHM of the individual polarisation fit increases as the LP approaches the QW resonance.}
\label{fig_S6} 
\end{center}
\end{figure}
The presence of the negative trion affects the bare exciton-polariton landscape considerably, which, in our case, is an undesirable situation. Characterisation of the sample using a PL excitation (PLE) measurement shown in Figure S.4 indicated that another excitation "window" occurred between 790 nm $\rightarrow$ 800 nm, although the signal generated in this window is significantly less intense for the same input laser power. PL spectra captured at an excitation wavelength of 800 nm (see Figure 2.a of the main text) show a clear avoided crossing with no signs of the negative trion state. While this is a positive result for non-resonant measurements of this polariton system, resonant measurements such as those conducted in the main body of this paper are still subject to interaction with the negative trion state. This is directly seen in the resonant LP lifetime behaviour for increasing cavity detuning shown in Figure 2.c of the main text. Following this approach, $\tau_{LP}$ can be expressed as:
	\begin{equation}
    \frac{1}{\tau_{LP}} = |C_X|^2 \gamma_{L} + \frac{1-|C_X|^2}{\tau_C}
    \label{eqHopfield}
	\end{equation}
where $C_X$ is the Hopfield coefficient and $\tau_C$ is the bare cavity lifetime. $\gamma_{L}$ corresponds to a loss rate for polaritons due to the elastic scattering towards the localized states. Mathematically, this behavior is best captured by a Gaussian distribution of the following form
	\begin{equation}
    \gamma_{L} = S\; e^{\frac{(E_{LP}-E_{X^0}-E_B)^2}{2 \sigma_{L}^2}}
    \label{eqGammaLDefn}
	\end{equation}
where S is the peak elastic scattering rate. $E_{LP}$ and $E_{X^0 }$ are the LP and exciton energies, and $E_B$ is the trion binding energy; $\sigma_{L}$  denotes the width of the Gaussian. Both $E_B$ and $\sigma_{L}$ can be extracted from photoluminescence spectra (see inset in figure S.5). Figure S.5 shows $\tau_{LP}$ data and the best fit to this model using the trion transition (red continuous line) and the bare exciton transition (black dashed line) as the origin of the scattering losses. As shown, the best model fit using $X^0$ transition (i.e. $E_B = 0$ and $\sigma = \sigma_X^0$) deviates substantially from the experimental trend.  

Finally, in order to include the effect of the polariton losses into the main master equation model, we define an effective excitonic lifetime as:
    \begin{equation}
    \frac{1}{\tau^{\text{eff}}_{X}} =\gamma_{L}.
	\end{equation}
Following this definition we include $\tau^{\text{eff}}_{X}$ as a parameter in the master equation model described above, identifying $\gamma_X = \gamma_L$, which encompasses all forms of excitonic loss. This effective parameter varies with $\Delta$, and is calculated via Eqs. (\ref{eqHopfield}) -- (\ref{eqGammaLDefn}) and the fitting shown in Figure S.5.

\subsection{Resonant polariton transmission spectrum}

Before correlation measurements were performed at each detuning as described in the main body of this paper, resonant polariton transmission spectra were taken using an M-Squared Cw laser locked to a High Finesse wavemeter (linewidth < 50kHz). This returns higher resolution spectra of the structure of the LP state compared to the Photoluminescence data. Figure S.6 shows the analysis of this transmission data. There is a clear polarisation splitting of the LP peak (50 - 90 $\mu$eV). This could be a result of imperfect symmetry of the machined indentation on the fibre component used in our fibre microcavities or the presence of birefringence in the mirror coatings and the GaAs itself. To extract the linewidth of the LP for increasing cavity-exciton detunings ($\Delta$), we have used two Lorentzian curves to describe the separate polarisations. The two curves were restricted to having the same area and linewidth. These restrictions are imposed to illustrate that each polarization has equal contributions to the LP structure. In Figure S.6 a,b,c,d the results of this analysis are shown with the two fitted Lorentzian curves shown in dash-dotted black lines, the total fitted curve in red and the data in black circles. Linewidth increases from 60 to 140 $\mu$eV from $\Delta =$ +0.07 to 0.82 meV. 

\subsection*{References}

\bibliography{main.bib}

\begin{thebibliography}{36}%
\makeatletter
\providecommand \@ifxundefined [1]{%
 \@ifx{#1\undefined}
}%
\providecommand \@ifnum [1]{%
 \ifnum #1\expandafter \@firstoftwo
 \else \expandafter \@secondoftwo
 \fi
}%
\providecommand \@ifx [1]{%
 \ifx #1\expandafter \@firstoftwo
 \else \expandafter \@secondoftwo
 \fi
}%
\providecommand \natexlab [1]{#1}%
\providecommand \enquote  [1]{``#1''}%
\providecommand \bibnamefont  [1]{#1}%
\providecommand \bibfnamefont [1]{#1}%
\providecommand \citenamefont [1]{#1}%
\providecommand \href@noop [0]{\@secondoftwo}%
\providecommand \href [0]{\begingroup \@sanitize@url \@href}%
\providecommand \@href[1]{\@@startlink{#1}\@@href}%
\providecommand \@@href[1]{\endgroup#1\@@endlink}%
\providecommand \@sanitize@url [0]{\catcode `\\12\catcode `\$12\catcode
  `\&12\catcode `\#12\catcode `\^12\catcode `\_12\catcode `\%12\relax}%
\providecommand \@@startlink[1]{}%
\providecommand \@@endlink[0]{}%
\providecommand \url  [0]{\begingroup\@sanitize@url \@url }%
\providecommand \@url [1]{\endgroup\@href {#1}{\urlprefix }}%
\providecommand \urlprefix  [0]{URL }%
\providecommand \Eprint [0]{\href }%
\providecommand \doibase [0]{http://dx.doi.org/}%
\providecommand \selectlanguage [0]{\@gobble}%
\providecommand \bibinfo  [0]{\@secondoftwo}%
\providecommand \bibfield  [0]{\@secondoftwo}%
\providecommand \translation [1]{[#1]}%
\providecommand \BibitemOpen [0]{}%
\providecommand \bibitemStop [0]{}%
\providecommand \bibitemNoStop [0]{.\EOS\space}%
\providecommand \EOS [0]{\spacefactor3000\relax}%
\providecommand \BibitemShut  [1]{\csname bibitem#1\endcsname}%
\let\auto@bib@innerbib\@empty
\bibitem [{\citenamefont {Carusotto}\ and\ \citenamefont
  {Ciuti}(2013)}]{Carusotto2013}%
  \BibitemOpen
  \bibfield  {author} {\bibinfo {author} {\bibfnamefont {I.}~\bibnamefont
  {Carusotto}}\ and\ \bibinfo {author} {\bibfnamefont {C.}~\bibnamefont
  {Ciuti}},\ }\href {\doibase 10.1103/RevModPhys.85.299} {\bibfield  {journal}
  {\bibinfo  {journal} {Rev. Mod. Phys.}\ }\textbf {\bibinfo {volume} {85}},\
  \bibinfo {pages} {299} (\bibinfo {year} {2013})}\BibitemShut {NoStop}%
\bibitem [{\citenamefont {Berloff}\ \emph {et~al.}(2017)\citenamefont
  {Berloff}, \citenamefont {Silva}, \citenamefont {Kalinin}, \citenamefont
  {Askitopoulos}, \citenamefont {T{\"{o}}pfer}, \citenamefont {Cilibrizzi},
  \citenamefont {Langbein},\ and\ \citenamefont {Lagoudakis}}]{Berloff2017}%
  \BibitemOpen
  \bibfield  {author} {\bibinfo {author} {\bibfnamefont {N.~G.}\ \bibnamefont
  {Berloff}}, \bibinfo {author} {\bibfnamefont {M.}~\bibnamefont {Silva}},
  \bibinfo {author} {\bibfnamefont {K.}~\bibnamefont {Kalinin}}, \bibinfo
  {author} {\bibfnamefont {A.}~\bibnamefont {Askitopoulos}}, \bibinfo {author}
  {\bibfnamefont {J.~D.}\ \bibnamefont {T{\"{o}}pfer}}, \bibinfo {author}
  {\bibfnamefont {P.}~\bibnamefont {Cilibrizzi}}, \bibinfo {author}
  {\bibfnamefont {W.}~\bibnamefont {Langbein}}, \ and\ \bibinfo {author}
  {\bibfnamefont {P.~G.}\ \bibnamefont {Lagoudakis}},\ }\href {\doibase
  http://dx.doi.org/10.1038/nmat4971} {\bibfield  {journal} {\bibinfo
  {journal} {Nat. Mater.}\ }\textbf {\bibinfo {volume} {16}},\ \bibinfo {pages}
  {1120} (\bibinfo {year} {2017})}\BibitemShut {NoStop}%
\bibitem [{\citenamefont {Jacqmin}\ \emph {et~al.}(2014)\citenamefont
  {Jacqmin}, \citenamefont {Carusotto}, \citenamefont {Sagnes}, \citenamefont
  {Abbarchi}, \citenamefont {Solnyshkov}, \citenamefont {Malpuech},
  \citenamefont {Galopin}, \citenamefont {Lema\^{\i}tre}, \citenamefont
  {Bloch},\ and\ \citenamefont {Amo}}]{Jacqmin2014}%
  \BibitemOpen
  \bibfield  {author} {\bibinfo {author} {\bibfnamefont {T.}~\bibnamefont
  {Jacqmin}}, \bibinfo {author} {\bibfnamefont {I.}~\bibnamefont {Carusotto}},
  \bibinfo {author} {\bibfnamefont {I.}~\bibnamefont {Sagnes}}, \bibinfo
  {author} {\bibfnamefont {M.}~\bibnamefont {Abbarchi}}, \bibinfo {author}
  {\bibfnamefont {D.~D.}\ \bibnamefont {Solnyshkov}}, \bibinfo {author}
  {\bibfnamefont {G.}~\bibnamefont {Malpuech}}, \bibinfo {author}
  {\bibfnamefont {E.}~\bibnamefont {Galopin}}, \bibinfo {author} {\bibfnamefont
  {A.}~\bibnamefont {Lema\^{\i}tre}}, \bibinfo {author} {\bibfnamefont
  {J.}~\bibnamefont {Bloch}}, \ and\ \bibinfo {author} {\bibfnamefont
  {A.}~\bibnamefont {Amo}},\ }\href {\doibase 10.1103/PhysRevLett.112.116402}
  {\bibfield  {journal} {\bibinfo  {journal} {Phys. Rev. Lett.}\ }\textbf
  {\bibinfo {volume} {112}},\ \bibinfo {pages} {116402} (\bibinfo {year}
  {2014})}\BibitemShut {NoStop}%
\bibitem [{\citenamefont {Baboux}\ \emph {et~al.}(2016)\citenamefont {Baboux},
  \citenamefont {Ge}, \citenamefont {Jacqmin}, \citenamefont {Biondi},
  \citenamefont {Galopin}, \citenamefont {Lema\^{\i}tre}, \citenamefont
  {Le~Gratiet}, \citenamefont {Sagnes}, \citenamefont {Schmidt}, \citenamefont
  {T\"ureci}, \citenamefont {Amo},\ and\ \citenamefont {Bloch}}]{Baboux2016}%
  \BibitemOpen
  \bibfield  {author} {\bibinfo {author} {\bibfnamefont {F.}~\bibnamefont
  {Baboux}}, \bibinfo {author} {\bibfnamefont {L.}~\bibnamefont {Ge}}, \bibinfo
  {author} {\bibfnamefont {T.}~\bibnamefont {Jacqmin}}, \bibinfo {author}
  {\bibfnamefont {M.}~\bibnamefont {Biondi}}, \bibinfo {author} {\bibfnamefont
  {E.}~\bibnamefont {Galopin}}, \bibinfo {author} {\bibfnamefont
  {A.}~\bibnamefont {Lema\^{\i}tre}}, \bibinfo {author} {\bibfnamefont
  {L.}~\bibnamefont {Le~Gratiet}}, \bibinfo {author} {\bibfnamefont
  {I.}~\bibnamefont {Sagnes}}, \bibinfo {author} {\bibfnamefont
  {S.}~\bibnamefont {Schmidt}}, \bibinfo {author} {\bibfnamefont {H.~E.}\
  \bibnamefont {T\"ureci}}, \bibinfo {author} {\bibfnamefont {A.}~\bibnamefont
  {Amo}}, \ and\ \bibinfo {author} {\bibfnamefont {J.}~\bibnamefont {Bloch}},\
  }\href {\doibase 10.1103/PhysRevLett.116.066402} {\bibfield  {journal}
  {\bibinfo  {journal} {Phys. Rev. Lett.}\ }\textbf {\bibinfo {volume} {116}},\
  \bibinfo {pages} {066402} (\bibinfo {year} {2016})}\BibitemShut {NoStop}%
\bibitem [{\citenamefont {Dagvadorj}\ \emph {et~al.}(2015)\citenamefont
  {Dagvadorj}, \citenamefont {Fellows}, \citenamefont
  {Matyja\ifmmode~\acute{s}\else \'{s}\fi{}kiewicz}, \citenamefont {Marchetti},
  \citenamefont {Carusotto},\ and\ \citenamefont {Szyma\ifmmode~\acute{n}\else
  \'{n}\fi{}ska}}]{Dagvadorj2015}%
  \BibitemOpen
  \bibfield  {author} {\bibinfo {author} {\bibfnamefont {G.}~\bibnamefont
  {Dagvadorj}}, \bibinfo {author} {\bibfnamefont {J.~M.}\ \bibnamefont
  {Fellows}}, \bibinfo {author} {\bibfnamefont {S.}~\bibnamefont
  {Matyja\ifmmode~\acute{s}\else \'{s}\fi{}kiewicz}}, \bibinfo {author}
  {\bibfnamefont {F.~M.}\ \bibnamefont {Marchetti}}, \bibinfo {author}
  {\bibfnamefont {I.}~\bibnamefont {Carusotto}}, \ and\ \bibinfo {author}
  {\bibfnamefont {M.~H.}\ \bibnamefont {Szyma\ifmmode~\acute{n}\else
  \'{n}\fi{}ska}},\ }\href {\doibase 10.1103/PhysRevX.5.041028} {\bibfield
  {journal} {\bibinfo  {journal} {Phys. Rev. X}\ }\textbf {\bibinfo {volume}
  {5}},\ \bibinfo {pages} {041028} (\bibinfo {year} {2015})}\BibitemShut
  {NoStop}%
\bibitem [{\citenamefont {Boulier}\ \emph {et~al.}(2014)\citenamefont
  {Boulier}, \citenamefont {Bamba}, \citenamefont {Amo}, \citenamefont
  {Adrados}, \citenamefont {Lema{\^{i}}tre}, \citenamefont {Galopin},
  \citenamefont {Sagnes}, \citenamefont {Bloch}, \citenamefont {Ciuti},
  \citenamefont {Giacobino},\ and\ \citenamefont {Bramati}}]{Bramati2014}%
  \BibitemOpen
  \bibfield  {author} {\bibinfo {author} {\bibfnamefont {T.}~\bibnamefont
  {Boulier}}, \bibinfo {author} {\bibfnamefont {M.}~\bibnamefont {Bamba}},
  \bibinfo {author} {\bibfnamefont {A.}~\bibnamefont {Amo}}, \bibinfo {author}
  {\bibfnamefont {C.}~\bibnamefont {Adrados}}, \bibinfo {author} {\bibfnamefont
  {A.}~\bibnamefont {Lema{\^{i}}tre}}, \bibinfo {author} {\bibfnamefont
  {E.}~\bibnamefont {Galopin}}, \bibinfo {author} {\bibfnamefont
  {I.}~\bibnamefont {Sagnes}}, \bibinfo {author} {\bibfnamefont
  {J.}~\bibnamefont {Bloch}}, \bibinfo {author} {\bibfnamefont
  {C.}~\bibnamefont {Ciuti}}, \bibinfo {author} {\bibfnamefont
  {E.}~\bibnamefont {Giacobino}}, \ and\ \bibinfo {author} {\bibfnamefont
  {A.}~\bibnamefont {Bramati}},\ }\href {\doibase
  http://dx.doi.org/10.1038/ncomms4260} {\bibfield  {journal} {\bibinfo
  {journal} {Nat. Commun.}\ }\textbf {\bibinfo {volume} {5}},\ \bibinfo {pages}
  {3260} (\bibinfo {year} {2014})}\BibitemShut {NoStop}%
\bibitem [{\citenamefont {Sanvitto}\ and\ \citenamefont
  {K{\'{e}}na-Cohen}(2016)}]{Sanvitto2016}%
  \BibitemOpen
  \bibfield  {author} {\bibinfo {author} {\bibfnamefont {D.}~\bibnamefont
  {Sanvitto}}\ and\ \bibinfo {author} {\bibfnamefont {S.}~\bibnamefont
  {K{\'{e}}na-Cohen}},\ }\href {\doibase http://dx.doi.org/10.1038/nmat4668}
  {\bibfield  {journal} {\bibinfo  {journal} {Nat. Mater.}\ }\textbf {\bibinfo
  {volume} {15}},\ \bibinfo {pages} {1061} (\bibinfo {year}
  {2016})}\BibitemShut {NoStop}%
\bibitem [{\citenamefont {Noh}\ and\ \citenamefont
  {Angelakis}(2017)}]{Angelakis2017}%
  \BibitemOpen
  \bibfield  {author} {\bibinfo {author} {\bibfnamefont {C.}~\bibnamefont
  {Noh}}\ and\ \bibinfo {author} {\bibfnamefont {D.~G.}\ \bibnamefont
  {Angelakis}},\ }\href@noop {} {\bibfield  {journal} {\bibinfo  {journal}
  {Rep. Prog. Phys.}\ }\textbf {\bibinfo {volume} {80}},\ \bibinfo {pages}
  {016401} (\bibinfo {year} {2017})}\BibitemShut {NoStop}%
\bibitem [{\citenamefont {Verger}\ \emph {et~al.}(2006)\citenamefont {Verger},
  \citenamefont {Ciuti},\ and\ \citenamefont {Carusotto}}]{Verger2006}%
  \BibitemOpen
  \bibfield  {author} {\bibinfo {author} {\bibfnamefont {A.}~\bibnamefont
  {Verger}}, \bibinfo {author} {\bibfnamefont {C.}~\bibnamefont {Ciuti}}, \
  and\ \bibinfo {author} {\bibfnamefont {I.}~\bibnamefont {Carusotto}},\ }\href
  {\doibase 10.1103/PhysRevB.73.193306} {\bibfield  {journal} {\bibinfo
  {journal} {Phys. Rev. B}\ }\textbf {\bibinfo {volume} {73}},\ \bibinfo
  {pages} {193306} (\bibinfo {year} {2006})}\BibitemShut {NoStop}%
\bibitem [{\citenamefont {Savasta}\ \emph {et~al.}(2005)\citenamefont
  {Savasta}, \citenamefont {Stefano}, \citenamefont {Savona},\ and\
  \citenamefont {Langbein}}]{Savasta2005}%
  \BibitemOpen
  \bibfield  {author} {\bibinfo {author} {\bibfnamefont {S.}~\bibnamefont
  {Savasta}}, \bibinfo {author} {\bibfnamefont {O.~D.}\ \bibnamefont
  {Stefano}}, \bibinfo {author} {\bibfnamefont {V.}~\bibnamefont {Savona}}, \
  and\ \bibinfo {author} {\bibfnamefont {W.}~\bibnamefont {Langbein}},\ }\href
  {\doibase 10.1103/PhysRevLett.94.246401} {\bibfield  {journal} {\bibinfo
  {journal} {Phys. Rev. Lett.}\ }\textbf {\bibinfo {volume} {94}},\ \bibinfo
  {pages} {246401} (\bibinfo {year} {2005})}\BibitemShut {NoStop}%
\bibitem [{\citenamefont {Amo}\ \emph {et~al.}(2010)\citenamefont {Amo},
  \citenamefont {Liew}, \citenamefont {Adrados}, \citenamefont {Houdre},
  \citenamefont {Giacobino}, \citenamefont {Kavokin},\ and\ \citenamefont
  {Bramati}}]{Amo2010}%
  \BibitemOpen
  \bibfield  {author} {\bibinfo {author} {\bibfnamefont {A.}~\bibnamefont
  {Amo}}, \bibinfo {author} {\bibfnamefont {T.}~\bibnamefont {Liew}}, \bibinfo
  {author} {\bibfnamefont {C.}~\bibnamefont {Adrados}}, \bibinfo {author}
  {\bibfnamefont {R.}~\bibnamefont {Houdre}}, \bibinfo {author} {\bibfnamefont
  {E.}~\bibnamefont {Giacobino}}, \bibinfo {author} {\bibfnamefont
  {A.}~\bibnamefont {Kavokin}}, \ and\ \bibinfo {author} {\bibfnamefont
  {A.}~\bibnamefont {Bramati}},\ }\href {\doibase 10.1038/nphoton.2010.79}
  {\bibfield  {journal} {\bibinfo  {journal} {Nature Photon.}\ }\textbf
  {\bibinfo {volume} {4}},\ \bibinfo {pages} {361} (\bibinfo {year}
  {2010})}\BibitemShut {NoStop}%
\bibitem [{\citenamefont {Imamo{\u{g}}lu}\ \emph {et~al.}(1997)\citenamefont
  {Imamo{\u{g}}lu}, \citenamefont {Schmidt}, \citenamefont {Woods},\ and\
  \citenamefont {Deutsch}}]{Imamoglu1997}%
  \BibitemOpen
  \bibfield  {author} {\bibinfo {author} {\bibfnamefont {A.}~\bibnamefont
  {Imamo{\u{g}}lu}}, \bibinfo {author} {\bibfnamefont {H.}~\bibnamefont
  {Schmidt}}, \bibinfo {author} {\bibfnamefont {G.}~\bibnamefont {Woods}}, \
  and\ \bibinfo {author} {\bibfnamefont {M.}~\bibnamefont {Deutsch}},\ }\href
  {\doibase 10.1103/PhysRevLett.79.1467} {\bibfield  {journal} {\bibinfo
  {journal} {Phys. Rev. Lett.}\ }\textbf {\bibinfo {volume} {79}},\ \bibinfo
  {pages} {1467} (\bibinfo {year} {1997})}\BibitemShut {NoStop}%
\bibitem [{\citenamefont {Liebisch}\ \emph {et~al.}(2005)\citenamefont
  {Liebisch}, \citenamefont {Reinhard}, \citenamefont {Berman},\ and\
  \citenamefont {Raithel}}]{Liebisch2005}%
  \BibitemOpen
  \bibfield  {author} {\bibinfo {author} {\bibfnamefont {T.~C.}\ \bibnamefont
  {Liebisch}}, \bibinfo {author} {\bibfnamefont {A.}~\bibnamefont {Reinhard}},
  \bibinfo {author} {\bibfnamefont {P.~R.}\ \bibnamefont {Berman}}, \ and\
  \bibinfo {author} {\bibfnamefont {G.}~\bibnamefont {Raithel}},\ }\href
  {\doibase 10.1103/PhysRevLett.95.253002} {\bibfield  {journal} {\bibinfo
  {journal} {Phys. Rev. Lett.}\ }\textbf {\bibinfo {volume} {95}},\ \bibinfo
  {pages} {253002} (\bibinfo {year} {2005})}\BibitemShut {NoStop}%
\bibitem [{\citenamefont {Jia}\ \emph {et~al.}(2017)\citenamefont {Jia},
  \citenamefont {Schine}, \citenamefont {Georgakopoulos}, \citenamefont {Ryou},
  \citenamefont {Sommer},\ and\ \citenamefont {Simon}}]{Ningyuan2017}%
  \BibitemOpen
  \bibfield  {author} {\bibinfo {author} {\bibfnamefont {N.}~\bibnamefont
  {Jia}}, \bibinfo {author} {\bibfnamefont {N.}~\bibnamefont {Schine}},
  \bibinfo {author} {\bibfnamefont {A.}~\bibnamefont {Georgakopoulos}},
  \bibinfo {author} {\bibfnamefont {A.}~\bibnamefont {Ryou}}, \bibinfo {author}
  {\bibfnamefont {A.}~\bibnamefont {Sommer}}, \ and\ \bibinfo {author}
  {\bibfnamefont {J.}~\bibnamefont {Simon}},\ }\href@noop {} {\bibfield
  {journal} {\bibinfo  {journal} {ArXiv}\ }\textbf {\bibinfo {volume} {--}},\
  \bibinfo {pages} {1705.07475} (\bibinfo {year} {2017})}\BibitemShut {NoStop}%
\bibitem [{\citenamefont {Carusotto}\ \emph {et~al.}(2010)\citenamefont
  {Carusotto}, \citenamefont {Volz},\ and\ \citenamefont
  {Imamo{\u{g}}lu}}]{Carusotto2010}%
  \BibitemOpen
  \bibfield  {author} {\bibinfo {author} {\bibfnamefont {I.}~\bibnamefont
  {Carusotto}}, \bibinfo {author} {\bibfnamefont {T.}~\bibnamefont {Volz}}, \
  and\ \bibinfo {author} {\bibfnamefont {A.}~\bibnamefont {Imamo{\u{g}}lu}},\
  }\href@noop {} {\bibfield  {journal} {\bibinfo  {journal} {EPL}\ }\textbf
  {\bibinfo {volume} {90}},\ \bibinfo {pages} {37001} (\bibinfo {year}
  {2010})}\BibitemShut {NoStop}%
\bibitem [{\citenamefont {Miguel-S{\'{a}}nchez}\ \emph
  {et~al.}(2013)\citenamefont {Miguel-S{\'{a}}nchez}, \citenamefont {Reinhard},
  \citenamefont {Togan}, \citenamefont {Volz}, \citenamefont {Imamo{\u{g}}lu},
  \citenamefont {Besga}, \citenamefont {Reichel},\ and\ \citenamefont
  {Est{\`{e}}ve}}]{Sanchez2013}%
  \BibitemOpen
  \bibfield  {author} {\bibinfo {author} {\bibfnamefont {J.}~\bibnamefont
  {Miguel-S{\'{a}}nchez}}, \bibinfo {author} {\bibfnamefont {A.}~\bibnamefont
  {Reinhard}}, \bibinfo {author} {\bibfnamefont {E.}~\bibnamefont {Togan}},
  \bibinfo {author} {\bibfnamefont {T.}~\bibnamefont {Volz}}, \bibinfo {author}
  {\bibfnamefont {A.}~\bibnamefont {Imamo{\u{g}}lu}}, \bibinfo {author}
  {\bibfnamefont {B.}~\bibnamefont {Besga}}, \bibinfo {author} {\bibfnamefont
  {J.}~\bibnamefont {Reichel}}, \ and\ \bibinfo {author} {\bibfnamefont
  {J.}~\bibnamefont {Est{\`{e}}ve}},\ }\href@noop {} {\bibfield  {journal}
  {\bibinfo  {journal} {New J. Phys.}\ }\textbf {\bibinfo {volume} {15}},\
  \bibinfo {pages} {045002} (\bibinfo {year} {2013})}\BibitemShut {NoStop}%
\bibitem [{\citenamefont {Reinhard}\ \emph {et~al.}(2012)\citenamefont
  {Reinhard}, \citenamefont {Volz}, \citenamefont {Winger}, \citenamefont
  {Badolato}, \citenamefont {Hennessy}, \citenamefont {Hu},\ and\ \citenamefont
  {Imamo{\u{g}}lu}}]{Reinhard2012}%
  \BibitemOpen
  \bibfield  {author} {\bibinfo {author} {\bibfnamefont {A.}~\bibnamefont
  {Reinhard}}, \bibinfo {author} {\bibfnamefont {T.}~\bibnamefont {Volz}},
  \bibinfo {author} {\bibfnamefont {M.}~\bibnamefont {Winger}}, \bibinfo
  {author} {\bibfnamefont {A.}~\bibnamefont {Badolato}}, \bibinfo {author}
  {\bibfnamefont {K.~J.}\ \bibnamefont {Hennessy}}, \bibinfo {author}
  {\bibfnamefont {E.~L.}\ \bibnamefont {Hu}}, \ and\ \bibinfo {author}
  {\bibfnamefont {A.}~\bibnamefont {Imamo{\u{g}}lu}},\ }\href {\doibase
  http://dx.doi.org/10.1038/nphoton.2011.321} {\bibfield  {journal} {\bibinfo
  {journal} {Nature Photon.}\ }\textbf {\bibinfo {volume} {6}},\ \bibinfo
  {pages} {93} (\bibinfo {year} {2012})}\BibitemShut {NoStop}%
\bibitem [{\citenamefont {Fink}\ \emph {et~al.}(2017)\citenamefont {Fink},
  \citenamefont {Schade}, \citenamefont {H{\"{o}fling}}, \citenamefont
  {Schneider},\ and\ \citenamefont {Imamo{\u{g}}lu}}]{Imamoglu2017}%
  \BibitemOpen
  \bibfield  {author} {\bibinfo {author} {\bibfnamefont {T.}~\bibnamefont
  {Fink}}, \bibinfo {author} {\bibfnamefont {A.}~\bibnamefont {Schade}},
  \bibinfo {author} {\bibfnamefont {S.}~\bibnamefont {H{\"{o}fling}}}, \bibinfo
  {author} {\bibfnamefont {C.}~\bibnamefont {Schneider}}, \ and\ \bibinfo
  {author} {\bibfnamefont {A.}~\bibnamefont {Imamo{\u{g}}lu}},\ }\href
  {\doibase 10.1038/s41567-017-0020-9} {\bibfield  {journal} {\bibinfo
  {journal} {Nat. Phys.}\ } (\bibinfo {year} {2017}),\
  10.1038/s41567-017-0020-9}\BibitemShut {NoStop}%
\bibitem [{\citenamefont {Deveaud}\ \emph {et~al.}(2005)\citenamefont
  {Deveaud}, \citenamefont {Kappei}, \citenamefont {Berney}, \citenamefont
  {Morier-Genoud}, \citenamefont {Portella-Oberli}, \citenamefont {Szczytko},\
  and\ \citenamefont {Piermarocchi}}]{Deveaud2005}%
  \BibitemOpen
  \bibfield  {author} {\bibinfo {author} {\bibfnamefont {B.}~\bibnamefont
  {Deveaud}}, \bibinfo {author} {\bibfnamefont {L.}~\bibnamefont {Kappei}},
  \bibinfo {author} {\bibfnamefont {J.}~\bibnamefont {Berney}}, \bibinfo
  {author} {\bibfnamefont {F.}~\bibnamefont {Morier-Genoud}}, \bibinfo {author}
  {\bibfnamefont {M.}~\bibnamefont {Portella-Oberli}}, \bibinfo {author}
  {\bibfnamefont {J.}~\bibnamefont {Szczytko}}, \ and\ \bibinfo {author}
  {\bibfnamefont {C.}~\bibnamefont {Piermarocchi}},\ }\href@noop {} {\bibfield
  {journal} {\bibinfo  {journal} {Chem. Phys.}\ }\textbf {\bibinfo {volume}
  {318}},\ \bibinfo {pages} {104} (\bibinfo {year} {2005})}\BibitemShut
  {NoStop}%
\bibitem [{\citenamefont {Mu{\~{n}}oz-Matutano}\ \emph
  {et~al.}(2008)\citenamefont {Mu{\~{n}}oz-Matutano}, \citenamefont
  {Al{\'{e}}n}, \citenamefont {Mart{\'{i}}nez-Pastor}, \citenamefont
  {Seravalli}, \citenamefont {Frigeri},\ and\ \citenamefont
  {Franchi}}]{MunozMatutano2008}%
  \BibitemOpen
  \bibfield  {author} {\bibinfo {author} {\bibfnamefont {G.}~\bibnamefont
  {Mu{\~{n}}oz-Matutano}}, \bibinfo {author} {\bibfnamefont {B.}~\bibnamefont
  {Al{\'{e}}n}}, \bibinfo {author} {\bibfnamefont {J.}~\bibnamefont
  {Mart{\'{i}}nez-Pastor}}, \bibinfo {author} {\bibfnamefont {L.}~\bibnamefont
  {Seravalli}}, \bibinfo {author} {\bibfnamefont {P.}~\bibnamefont {Frigeri}},
  \ and\ \bibinfo {author} {\bibfnamefont {S.}~\bibnamefont {Franchi}},\
  }\href@noop {} {\bibfield  {journal} {\bibinfo  {journal} {Nanotechnology}\
  }\textbf {\bibinfo {volume} {19}},\ \bibinfo {pages} {145711} (\bibinfo
  {year} {2008})}\BibitemShut {NoStop}%
\bibitem [{\citenamefont {Sermage}\ \emph {et~al.}(1996)\citenamefont
  {Sermage}, \citenamefont {Long}, \citenamefont {Abram}, \citenamefont
  {Marzin}, \citenamefont {Bloch}, \citenamefont {Planel},\ and\ \citenamefont
  {Thierry-Mieg}}]{Sermage1996}%
  \BibitemOpen
  \bibfield  {author} {\bibinfo {author} {\bibfnamefont {B.}~\bibnamefont
  {Sermage}}, \bibinfo {author} {\bibfnamefont {S.}~\bibnamefont {Long}},
  \bibinfo {author} {\bibfnamefont {I.}~\bibnamefont {Abram}}, \bibinfo
  {author} {\bibfnamefont {J.~Y.}\ \bibnamefont {Marzin}}, \bibinfo {author}
  {\bibfnamefont {J.}~\bibnamefont {Bloch}}, \bibinfo {author} {\bibfnamefont
  {R.}~\bibnamefont {Planel}}, \ and\ \bibinfo {author} {\bibfnamefont
  {V.}~\bibnamefont {Thierry-Mieg}},\ }\href@noop {} {\bibfield  {journal}
  {\bibinfo  {journal} {Phys. Rev. B}\ }\textbf {\bibinfo {volume} {53}},\
  \bibinfo {pages} {16516} (\bibinfo {year} {1996})}\BibitemShut {NoStop}%
\bibitem [{\citenamefont {Klembt}\ \emph {et~al.}(2017)\citenamefont {Klembt},
  \citenamefont {Stepanov}, \citenamefont {Klein}, \citenamefont {Minguzzi},\
  and\ \citenamefont {Richard}}]{Klembt2017}%
  \BibitemOpen
  \bibfield  {author} {\bibinfo {author} {\bibfnamefont {S.}~\bibnamefont
  {Klembt}}, \bibinfo {author} {\bibfnamefont {P.}~\bibnamefont {Stepanov}},
  \bibinfo {author} {\bibfnamefont {T.}~\bibnamefont {Klein}}, \bibinfo
  {author} {\bibfnamefont {A.}~\bibnamefont {Minguzzi}}, \ and\ \bibinfo
  {author} {\bibfnamefont {M.}~\bibnamefont {Richard}},\ }\href@noop {}
  {\bibfield  {journal} {\bibinfo  {journal} {ArXiv}\ ,\ \bibinfo {pages}
  {1603.04206}} (\bibinfo {year} {2017})}\BibitemShut {NoStop}%
\bibitem [{\citenamefont {Besga}\ \emph {et~al.}(2015)\citenamefont {Besga},
  \citenamefont {Vaneph}, \citenamefont {Reichel}, \citenamefont
  {Est{\`{e}}ve}, \citenamefont {Reinhard}, \citenamefont
  {Miguel-S{\'{a}}nchez}, \citenamefont {Imamo{\u{g}}lu},\ and\ \citenamefont
  {Volz}}]{Besga2015}%
  \BibitemOpen
  \bibfield  {author} {\bibinfo {author} {\bibfnamefont {B.}~\bibnamefont
  {Besga}}, \bibinfo {author} {\bibfnamefont {C.}~\bibnamefont {Vaneph}},
  \bibinfo {author} {\bibfnamefont {J.}~\bibnamefont {Reichel}}, \bibinfo
  {author} {\bibfnamefont {J.}~\bibnamefont {Est{\`{e}}ve}}, \bibinfo {author}
  {\bibfnamefont {A.}~\bibnamefont {Reinhard}}, \bibinfo {author}
  {\bibfnamefont {J.}~\bibnamefont {Miguel-S{\'{a}}nchez}}, \bibinfo {author}
  {\bibfnamefont {A.}~\bibnamefont {Imamo{\u{g}}lu}}, \ and\ \bibinfo {author}
  {\bibfnamefont {T.}~\bibnamefont {Volz}},\ }\href {\doibase
  10.1103/PhysRevApplied.3.014008} {\bibfield  {journal} {\bibinfo  {journal}
  {Phys. Rev. Appl.}\ }\textbf {\bibinfo {volume} {3}},\ \bibinfo {pages}
  {014008} (\bibinfo {year} {2015})}\BibitemShut {NoStop}%
\bibitem [{\citenamefont {Reitzenstein}\ \emph {et~al.}(2007)\citenamefont
  {Reitzenstein}, \citenamefont {Hofmann}, \citenamefont {Gorbunov},
  \citenamefont {Strau{\ss}}, \citenamefont {Kwon}, \citenamefont {Schneider},
  \citenamefont {L{\"{o}}ffler}, \citenamefont {H{\"{o}}fling}, \citenamefont
  {Kamp},\ and\ \citenamefont {Forchel}}]{Reitzenstein2007}%
  \BibitemOpen
  \bibfield  {author} {\bibinfo {author} {\bibfnamefont {S.}~\bibnamefont
  {Reitzenstein}}, \bibinfo {author} {\bibfnamefont {C.}~\bibnamefont
  {Hofmann}}, \bibinfo {author} {\bibfnamefont {A.}~\bibnamefont {Gorbunov}},
  \bibinfo {author} {\bibfnamefont {M.}~\bibnamefont {Strau{\ss}}}, \bibinfo
  {author} {\bibfnamefont {S.~H.}\ \bibnamefont {Kwon}}, \bibinfo {author}
  {\bibfnamefont {C.}~\bibnamefont {Schneider}}, \bibinfo {author}
  {\bibfnamefont {A.}~\bibnamefont {L{\"{o}}ffler}}, \bibinfo {author}
  {\bibfnamefont {S.}~\bibnamefont {H{\"{o}}fling}}, \bibinfo {author}
  {\bibfnamefont {M.}~\bibnamefont {Kamp}}, \ and\ \bibinfo {author}
  {\bibfnamefont {A.}~\bibnamefont {Forchel}},\ }\href@noop {} {\bibfield
  {journal} {\bibinfo  {journal} {Appl. Phys. Lett.}\ }\textbf {\bibinfo
  {volume} {90}},\ \bibinfo {pages} {251109} (\bibinfo {year}
  {2007})}\BibitemShut {NoStop}%
\bibitem [{\citenamefont {Wood}\ \emph {et~al.}(2018)\citenamefont {Wood},
  \citenamefont {Vidal}, \citenamefont {Mu{\~{n}}oz-Matutano},\ and\
  \citenamefont {Volz}}]{Wood2018}%
  \BibitemOpen
  \bibfield  {author} {\bibinfo {author} {\bibfnamefont {A.}~\bibnamefont
  {Wood}}, \bibinfo {author} {\bibfnamefont {X.}~\bibnamefont {Vidal}},
  \bibinfo {author} {\bibfnamefont {G.}~\bibnamefont {Mu{\~{n}}oz-Matutano}}, \
  and\ \bibinfo {author} {\bibfnamefont {T.}~\bibnamefont {Volz}},\ }\href@noop
  {} {\bibfield  {journal} {\bibinfo  {journal} {Arxiv}\ } (\bibinfo {year}
  {2018})}\BibitemShut {NoStop}%
\bibitem [{\citenamefont {Ciuti}\ \emph {et~al.}(2003)\citenamefont {Ciuti},
  \citenamefont {Schwendimann},\ and\ \citenamefont {Quattropani}}]{Ciuti2003}%
  \BibitemOpen
  \bibfield  {author} {\bibinfo {author} {\bibfnamefont {C.}~\bibnamefont
  {Ciuti}}, \bibinfo {author} {\bibfnamefont {P.}~\bibnamefont {Schwendimann}},
  \ and\ \bibinfo {author} {\bibfnamefont {A.}~\bibnamefont {Quattropani}},\
  }\href@noop {} {\bibfield  {journal} {\bibinfo  {journal} {Semicond. Sci.
  Technol}\ }\textbf {\bibinfo {volume} {18}},\ \bibinfo {pages} {S279}
  (\bibinfo {year} {2003})}\BibitemShut {NoStop}%
\bibitem [{\citenamefont {Ciuti}\ \emph {et~al.}(1998)\citenamefont {Ciuti},
  \citenamefont {Savona}, \citenamefont {Piermarocchi}, \citenamefont
  {Quattropani},\ and\ \citenamefont {Schwendimann}}]{Ciuti1998}%
  \BibitemOpen
  \bibfield  {author} {\bibinfo {author} {\bibfnamefont {C.}~\bibnamefont
  {Ciuti}}, \bibinfo {author} {\bibfnamefont {V.}~\bibnamefont {Savona}},
  \bibinfo {author} {\bibfnamefont {C.}~\bibnamefont {Piermarocchi}}, \bibinfo
  {author} {\bibfnamefont {A.}~\bibnamefont {Quattropani}}, \ and\ \bibinfo
  {author} {\bibfnamefont {P.}~\bibnamefont {Schwendimann}},\ }\href@noop {}
  {\bibfield  {journal} {\bibinfo  {journal} {Phys. Rev. B}\ }\textbf {\bibinfo
  {volume} {58}},\ \bibinfo {pages} {7926} (\bibinfo {year}
  {1998})}\BibitemShut {NoStop}%
\bibitem [{\citenamefont {Amo}\ \emph {et~al.}(2009)\citenamefont {Amo},
  \citenamefont {Lefr{\`{e}}re}, \citenamefont {Pigeon}, \citenamefont
  {Adrados}, \citenamefont {Ciuti}, \citenamefont {Carusotto}, \citenamefont
  {Houdr{\'{e}}}, \citenamefont {Giacobino},\ and\ \citenamefont
  {Bramati}}]{Amo2009}%
  \BibitemOpen
  \bibfield  {author} {\bibinfo {author} {\bibfnamefont {A.}~\bibnamefont
  {Amo}}, \bibinfo {author} {\bibfnamefont {J.}~\bibnamefont {Lefr{\`{e}}re}},
  \bibinfo {author} {\bibfnamefont {S.}~\bibnamefont {Pigeon}}, \bibinfo
  {author} {\bibfnamefont {C.}~\bibnamefont {Adrados}}, \bibinfo {author}
  {\bibfnamefont {C.}~\bibnamefont {Ciuti}}, \bibinfo {author} {\bibfnamefont
  {I.}~\bibnamefont {Carusotto}}, \bibinfo {author} {\bibfnamefont
  {R.}~\bibnamefont {Houdr{\'{e}}}}, \bibinfo {author} {\bibfnamefont
  {E.}~\bibnamefont {Giacobino}}, \ and\ \bibinfo {author} {\bibfnamefont
  {A.}~\bibnamefont {Bramati}},\ }\href@noop {} {\bibfield  {journal} {\bibinfo
   {journal} {Nat. Phys.}\ }\textbf {\bibinfo {volume} {5}},\ \bibinfo {pages}
  {805} (\bibinfo {year} {2009})}\BibitemShut {NoStop}%
\bibitem [{\citenamefont {Ferrier}\ \emph {et~al.}(2011)\citenamefont
  {Ferrier}, \citenamefont {Wertz}, \citenamefont {Johne}, \citenamefont
  {Solnyshkov}, \citenamefont {Senellart}, \citenamefont {Sagnes},
  \citenamefont {Lema\^{\i}tre}, \citenamefont {Malpuech},\ and\ \citenamefont
  {Bloch}}]{Ferrier2011}%
  \BibitemOpen
  \bibfield  {author} {\bibinfo {author} {\bibfnamefont {L.}~\bibnamefont
  {Ferrier}}, \bibinfo {author} {\bibfnamefont {E.}~\bibnamefont {Wertz}},
  \bibinfo {author} {\bibfnamefont {R.}~\bibnamefont {Johne}}, \bibinfo
  {author} {\bibfnamefont {D.~D.}\ \bibnamefont {Solnyshkov}}, \bibinfo
  {author} {\bibfnamefont {P.}~\bibnamefont {Senellart}}, \bibinfo {author}
  {\bibfnamefont {I.}~\bibnamefont {Sagnes}}, \bibinfo {author} {\bibfnamefont
  {A.}~\bibnamefont {Lema\^{\i}tre}}, \bibinfo {author} {\bibfnamefont
  {G.}~\bibnamefont {Malpuech}}, \ and\ \bibinfo {author} {\bibfnamefont
  {J.}~\bibnamefont {Bloch}},\ }\href {\doibase 10.1103/PhysRevLett.106.126401}
  {\bibfield  {journal} {\bibinfo  {journal} {Phys. Rev. Lett.}\ }\textbf
  {\bibinfo {volume} {106}},\ \bibinfo {pages} {126401} (\bibinfo {year}
  {2011})}\BibitemShut {NoStop}%
\bibitem [{\citenamefont {Dufferwiel}\ \emph {et~al.}(2014)\citenamefont
  {Dufferwiel}, \citenamefont {Fras}, \citenamefont {Trichet}, \citenamefont
  {Walker}, \citenamefont {Li}, \citenamefont {Giriunas}, \citenamefont
  {Makhonin}, \citenamefont {Wilson}, \citenamefont {Smith}, \citenamefont
  {Clarke}, \citenamefont {Skolnick},\ and\ \citenamefont
  {Krizhanovskii}}]{Dufferwiel2014}%
  \BibitemOpen
  \bibfield  {author} {\bibinfo {author} {\bibfnamefont {S.}~\bibnamefont
  {Dufferwiel}}, \bibinfo {author} {\bibfnamefont {F.}~\bibnamefont {Fras}},
  \bibinfo {author} {\bibfnamefont {A.}~\bibnamefont {Trichet}}, \bibinfo
  {author} {\bibfnamefont {P.~M.}\ \bibnamefont {Walker}}, \bibinfo {author}
  {\bibfnamefont {F.}~\bibnamefont {Li}}, \bibinfo {author} {\bibfnamefont
  {L.}~\bibnamefont {Giriunas}}, \bibinfo {author} {\bibfnamefont {M.~N.}\
  \bibnamefont {Makhonin}}, \bibinfo {author} {\bibfnamefont {L.~R.}\
  \bibnamefont {Wilson}}, \bibinfo {author} {\bibfnamefont {J.~M.}\
  \bibnamefont {Smith}}, \bibinfo {author} {\bibfnamefont {E.}~\bibnamefont
  {Clarke}}, \bibinfo {author} {\bibfnamefont {M.~S.}\ \bibnamefont
  {Skolnick}}, \ and\ \bibinfo {author} {\bibfnamefont {D.~N.}\ \bibnamefont
  {Krizhanovskii}},\ }\href {\doibase doi.org/10.1063/1.4878504} {\bibfield
  {journal} {\bibinfo  {journal} {Appl. Phys. Lett.}\ }\textbf {\bibinfo
  {volume} {104}},\ \bibinfo {pages} {192107} (\bibinfo {year}
  {2014})}\BibitemShut {NoStop}%
\bibitem [{\citenamefont {Richard}(2015)}]{Richard2015}%
  \BibitemOpen
  \bibfield  {author} {\bibinfo {author} {\bibfnamefont {M.}~\bibnamefont
  {Richard}},\ }\href@noop {} {\enquote {\bibinfo {title}
  {Light-excitons-phonons interaction in semiconductor nanostructures: from
  thermodynamics to optomechanics},}\ } (\bibinfo {year} {2015}),\ \bibinfo
  {note} {{A} dissertation submitted for the degree of Habilitation a diriger
  des recherches. Grenoble Universit{\'{e}}s, Institut N{\'{e}}el
  (CNRS)}\BibitemShut {NoStop}%
\bibitem [{\citenamefont {Yamamoto}\ and\ \citenamefont
  {Imamo{\u{g}}lu}(1999)}]{Yamamoto1999}%
  \BibitemOpen
  \bibfield  {author} {\bibinfo {author} {\bibfnamefont {Y.}~\bibnamefont
  {Yamamoto}}\ and\ \bibinfo {author} {\bibfnamefont {A.}~\bibnamefont
  {Imamo{\u{g}}lu}},\ }\href@noop {} {\emph {\bibinfo {title} {Mesoscopic
  Quantum Optics}}}\ (\bibinfo  {publisher} {John Wiley \& Sons Inc},\ \bibinfo
  {address} {New Jersey USA},\ \bibinfo {year} {1999})\BibitemShut {NoStop}%
\bibitem [{\citenamefont {Takemura}\ \emph {et~al.}(2014)\citenamefont
  {Takemura}, \citenamefont {Trebaol}, \citenamefont {Wouters}, \citenamefont
  {Portella-Oberli},\ and\ \citenamefont {Deveaud}}]{Deveaud2014}%
  \BibitemOpen
  \bibfield  {author} {\bibinfo {author} {\bibfnamefont {N.}~\bibnamefont
  {Takemura}}, \bibinfo {author} {\bibfnamefont {S.}~\bibnamefont {Trebaol}},
  \bibinfo {author} {\bibfnamefont {M.}~\bibnamefont {Wouters}}, \bibinfo
  {author} {\bibfnamefont {M.~T.}\ \bibnamefont {Portella-Oberli}}, \ and\
  \bibinfo {author} {\bibfnamefont {B.}~\bibnamefont {Deveaud}},\ }\href@noop
  {} {\bibfield  {journal} {\bibinfo  {journal} {Nat. Phys.}\ }\textbf
  {\bibinfo {volume} {10}},\ \bibinfo {pages} {500} (\bibinfo {year}
  {2014})}\BibitemShut {NoStop}%
\bibitem [{\citenamefont {Flayac}\ and\ \citenamefont
  {Savona}(2017)}]{Flayac2017}%
  \BibitemOpen
  \bibfield  {author} {\bibinfo {author} {\bibfnamefont {H.}~\bibnamefont
  {Flayac}}\ and\ \bibinfo {author} {\bibfnamefont {V.}~\bibnamefont
  {Savona}},\ }\href {\doibase 10.1103/PhysRevA.96.053810} {\bibfield
  {journal} {\bibinfo  {journal} {Phys. Rev. A}\ }\textbf {\bibinfo {volume}
  {96}},\ \bibinfo {pages} {053810} (\bibinfo {year} {2017})}\BibitemShut
  {NoStop}%
\bibitem [{\citenamefont {Low}\ \emph {et~al.}(2017)\citenamefont {Low},
  \citenamefont {Caldwell}, \citenamefont {Kumar}, \citenamefont {Fang},
  \citenamefont {Avouris}, \citenamefont {Heinz}, \citenamefont {Guinea},
  \citenamefont {Mart{\'{i}}n-Moreno},\ and\ \citenamefont
  {Koppens}}]{Low2017}%
  \BibitemOpen
  \bibfield  {author} {\bibinfo {author} {\bibfnamefont {T.}~\bibnamefont
  {Low}}, \bibinfo {author} {\bibfnamefont {J.~D.}\ \bibnamefont {Caldwell}},
  \bibinfo {author} {\bibfnamefont {A.}~\bibnamefont {Kumar}}, \bibinfo
  {author} {\bibfnamefont {N.~X.}\ \bibnamefont {Fang}}, \bibinfo {author}
  {\bibfnamefont {P.}~\bibnamefont {Avouris}}, \bibinfo {author} {\bibfnamefont
  {T.~F.}\ \bibnamefont {Heinz}}, \bibinfo {author} {\bibfnamefont
  {F.}~\bibnamefont {Guinea}}, \bibinfo {author} {\bibfnamefont
  {L.}~\bibnamefont {Mart{\'{i}}n-Moreno}}, \ and\ \bibinfo {author}
  {\bibfnamefont {F.}~\bibnamefont {Koppens}},\ }\href@noop {} {\bibfield
  {journal} {\bibinfo  {journal} {Nat. Mater.}\ }\textbf {\bibinfo {volume}
  {16}},\ \bibinfo {pages} {182} (\bibinfo {year} {2017})}\BibitemShut
  {NoStop}%
\bibitem [{\citenamefont {Dufferwiel}\ \emph {et~al.}(2017)\citenamefont
  {Dufferwiel}, \citenamefont {Lyons}, \citenamefont {Solnyshkov},
  \citenamefont {Trichet}, \citenamefont {Withers}, \citenamefont {Schwarz},
  \citenamefont {Malpuech}, \citenamefont {Smith}, \citenamefont {Novoselov},
  \citenamefont {Skolnick}, \citenamefont {Krizhanovskii},\ and\ \citenamefont
  {Tartakovskii}}]{Dufferwiel2017}%
  \BibitemOpen
  \bibfield  {author} {\bibinfo {author} {\bibfnamefont {S.}~\bibnamefont
  {Dufferwiel}}, \bibinfo {author} {\bibfnamefont {T.~P.}\ \bibnamefont
  {Lyons}}, \bibinfo {author} {\bibfnamefont {D.~D.}\ \bibnamefont
  {Solnyshkov}}, \bibinfo {author} {\bibfnamefont {A.~A.~P.}\ \bibnamefont
  {Trichet}}, \bibinfo {author} {\bibfnamefont {F.}~\bibnamefont {Withers}},
  \bibinfo {author} {\bibfnamefont {S.}~\bibnamefont {Schwarz}}, \bibinfo
  {author} {\bibfnamefont {G.}~\bibnamefont {Malpuech}}, \bibinfo {author}
  {\bibfnamefont {J.~M.}\ \bibnamefont {Smith}}, \bibinfo {author}
  {\bibfnamefont {K.~S.}\ \bibnamefont {Novoselov}}, \bibinfo {author}
  {\bibfnamefont {M.~S.}\ \bibnamefont {Skolnick}}, \bibinfo {author}
  {\bibfnamefont {D.~N.}\ \bibnamefont {Krizhanovskii}}, \ and\ \bibinfo
  {author} {\bibfnamefont {A.~I.}\ \bibnamefont {Tartakovskii}},\ }\href
  {\doibase http://dx.doi.org/10.1038/nphoton.2017.125} {\bibfield  {journal}
  {\bibinfo  {journal} {Nature Photon.}\ }\textbf {\bibinfo {volume} {11}},\
  \bibinfo {pages} {497} (\bibinfo {year} {2017})}\BibitemShut {NoStop}%
\end{thebibliography}%








\end{document}